\documentclass{jfm}
\usepackage{graphicx}
\usepackage{amsmath}
\usepackage{overpic}
\usepackage{longtable}
\usepackage{epstopdf,epsfig}
\usepackage{color}
\usepackage{tabularx}
\usepackage{amssymb}
\usepackage{caption}
\usepackage[colorlinks,citecolor = blue, linkcolor=black, hyperindex, CJKbookmarks]{hyperref}

\shorttitle{From zonal flow to convection rolls in RB convection with free-slip plates}
\shortauthor{Q. Wang et al.}

\title{From zonal flow to convection rolls in Rayleigh-B\'enard convection with free-slip plates}

\author{Qi Wang\aff{1,2},
Kai Leong Chong\aff{1},
Richard J. A. M. Stevens\aff{1},
Roberto Verzicco\aff{3,4,1},
and
Detlef Lohse\aff{1,5}\corresp{\email{d.lohse@utwente.nl}},
}
\affiliation{\aff{1}Physics of Fluids Group and Max Planck Center Twente for Complex Fluid Dynamics, MESA+ Institute and J. M. Burgers Centre for Fluid Dynamics, University of Twente, Enschede, The Netherlands.
\aff{2}Department of Modern Mechanics, University of Science and Technology of China, Hefei 230027, China
\aff{3}Dipartimento di Ingegneria Industriale, University of Rome ``Tor Vergata”, Via del Politecnico 1, Roma 00133, Italy
\aff{4}Gran Sasso Science Institute - Viale F. Crispi, 767100 L'Aquila, Italy.
\aff{5}Max Planck Institute for Dynamics and Self-Organization, 37077 G$\ddot{\rm{o}}$ttingen, Germany

}

\begin{document}

\maketitle

\begin{abstract}
Rayleigh-B\'enard (RB) convection with free-slip plates and horizontally periodic boundary conditions is investigated using direct numerical simulations. Two configurations are considered, one is 
two-dimension (2D) RB convection and the other one 
 three-dimension (3D) RB convection with a rotating axis parallel to the plate, which for strong rotation mimics 2D RB convection. 

For the 2D simulations,  we 
explore the parameter range 
of  Rayleigh numbers $Ra$ from $10^7$ to $10^9$ and Prandtl numbers $Pr$ from $1$ to $100$. 
The effect of the width-to-height aspect ratio $\Gamma$ is investigated for $1\leq\Gamma\leq128$. We show that 
 zonal flow, which was observed, for example, by Goluskin \emph{et al}. \emph{J. Fluid. Mech.} 759, 360-385 (2014) for $\Gamma=2$, is only stable when $\Gamma$ is smaller than a critical value, which depends on $Ra$ and $Pr$.
  With increasing $\Gamma$, we  find a second 
 regime in which both zonal flow and different convection roll states can be statistically stable.
 For even larger $\Gamma$, in a third regime, 
 only convection roll states are statistically stable and zonal flow is not sustained. How many convection rolls form (or in other words, what the mean aspect ratio of an  individual roll is), depends on the initial conditions and on $Ra$ and $Pr$. For instance,
 for $Ra=10^8$ and  $Pr=10$, 
  the aspect ratio $\Gamma_r$ of an individual,  statistically stable convection roll can vary in a large range 
  between $16/11$ and $64$. A convection roll with an as large aspect ratio of $\Gamma_r = 64$ or more generally already with $\Gamma_r \gg 10$ can be seen as “localized” zonal flow, and indeed carries over various properties of the global zonal flow.  
  
  For the 3D simulations, we fix $Ra=10^7$ and $Pr=0.71$, and compare  the flow for 
  $\Gamma=8$ and $\Gamma = 16$. We first show that with increasing rotation rate both the flow structures and global quantities like the Nusselt number $Nu$ and the Reynolds number $Re$ increasingly behave like in the 2D case. We then demonstrate that with increasing aspect ratio $\Gamma$,  zonal flow, which was observed  for small $\Gamma=2\pi$ by von Hardenberg  \emph{et al}. \emph{Phys. Rev. Lett.} 15, 134501 (2015),
   completely disappears for $\Gamma=16$. 
   For such large $\Gamma$
    only convection roll states are statistically stable. In between, here for medium aspect ratio
     $\Gamma = 8$, the convection roll state and the zonal flow state are both statistically stable. What state is taken
     depends on the initial conditions, similarly as we found for the 2D case.

\end{abstract}


\section{Introduction}
Large scale so-called zonal flows, which display strong horizontal winds, can be observed in many buoyancy-driven flows. Typical examples include zonal flow in the atmosphere of Jupiter \citep{heimpel2005simulation,kaspi2018jupiter,kong2018origin} and other three Jovian planets \citep{ingersoll1990atmospheric,sun1993banded,cho1996morphogenesis,yano2003outer}, in the oceans \citep{maximenko2005observational,richards2006zonal,nadiga2006zonal}, and possibly in the Earth's outer core \citep{miyagoshi2010zonal}. In toroidal tokamak devices, zonal flows in the poloidal direction are crucial to confine plasmas magnetically \citep{diamond2005zonal}. 

How to study such flows? In general, Rayleigh-B\'enard (RB) convection \citep{ahlers2009heat,lohse2010small,chilla2012new}, i.e., a fluid in a container heated from below and cooled from above, is the paradigmatic model system for buoyancy-driven flows. The key control parameters are the Rayleigh number $Ra=g\alpha \Delta H^3/\nu \kappa$ and the Prandtl number $Pr=\nu/\kappa$. Here, $g$ is the gravitational acceleration, $\alpha$ the thermal expansion coefficient, $H$ the height of the fluid sample, $\Delta=T_b-T_t$ the temperature difference between the hot bottom and the cold top plate, $\kappa$ the thermal diffusivity, and $\nu$ the kinematic viscosity of the fluid. The third control parameter is the aspect ratio $\Gamma$, which is defined as the ratio of the width to the height of the container. The response of the system is characterized by the Nusselt number $Nu={QH}/{(k\Delta})$ and the Reynolds number $Re={UH}/{\nu}$, which indicate the non-dimensional heat transport and flow strength in the system, respectively.
Here $Q$ is the heat flux crossing the system and $U=\sqrt{\left<\boldsymbol{u}\cdot\boldsymbol{u}\right>_{V,t}}$ the characteristic velocity, where $\left<\right>_{V,t}$ indicates volume and time averaging. Indeed, to study zonal flow, RB convection with free-slip plates and horizontally periodic boundary conditions has commonly served as model system \citep{goluskin2014convectively,van2014effect,von2015generation,novi2019rapidly}. Here the free-slip at the plates is very
crucial to enable the zonal flow; for no-slip boundary conditions, zonal flow is significantly suppressed and it only exists  for very small $\Gamma$ \citep{van2014effect}. 

In the 2D version of the RB system with free-slip plates and horizontally periodic boundary conditions, indeed, for small $\Gamma=2$, zonal flow develops readily since the horizontal periodicity allows for a horizontal mean flow, while the free-slip boundaries apply no shear stresses to slow down the fluid. In addition, the two-dimensionality precludes transverse perturbations that could
 disrupt the mean flow \citep{van2014effect,goluskin2014convectively}. Such zonal flow in 2D RB convection has attracted quite some attention because of its relevance to thermal convection in the atmosphere \citep{seychelles2008thermal,seychelles2010intermittent,bouchet2012statistical}. 
 For  free-slip boundary conditions at the plates, 
 $Pr=1$,  $Ra\ge10^7$, and a small $\Gamma=2$,
 \cite{van2014effect} 
  found that a flow topology consisting of two shear layers with a predominant horizontal motion is formed.
  The flow in the lower half of the domain moves in the opposite direction as that in the top part. Most of the time, the heat transfer of this flow is $Nu\approx1$, while there are intermittent bursts in which $Nu\gg1$. 
  \cite{goluskin2014convectively} studied the characteristics of such 2D zonal flows in a periodic $\Gamma=2$ cell for 
  an extended parameter range 
  $10^3 \le Ra\le10^{10}$ and $1\le Pr \le 10$. They found that for $Pr\le2$, the zonal flow undergoes strong global oscillations on long timescales.  Also intermittent bursts in the heat transport as  in \cite{van2014effect} are observed. For $Pr\ge3$, the zonal flow is sustained at all times without bursts, and the Nusselt number $Nu$ is always much larger than 1.

To what degree can 2D simulations mimic the dynamics in three-dimensional (3D) flows? Actually many 3D geophysical and astrophysical flows exhibit certain 2D properties when  anisotropic effects, such as geometrical confinement, rapid rotation, stratification, or magnetic fields, are imposed.  We will show in this study how the 2D flow structures arise with increasing rotation rate for RB convection rotating about a horizontal axis.
Such flow will be called span-wise rotating RB convection in this paper.  2D simulations, which are computationally more accessible than 3D simulations, have also been widely used to study RB convection with no-slip plates \citep{johnston2009comparison,sugiyama2010flow,huang2013counter,van2015logarithmic,zhu2018transition,zhu2019nu,wan2020non}. In \cite{van2013comparison} 2D and 3D simulations are compared in detail, and many similarities are found for $Pr\ge1$.

In contrast to in 2D configuration, zonal flow has not been reported in horizontally isotropic 3D simulations of RB convection with free-slip plates \citep{petschel2013dissipation,kunnen2016transition}. It seems that in 3D convection zonal flow only appears when an anisotropy is added. For example, \cite{von2015generation} found that a strong zero-wavenumber wind (i.e., zonal flow) can arise in 3D RB convection if the horizontal isotropy is broken by strong enough
uniform rotation about a horizontal axis. Indeed, according to the Taylor-Proudman theorem, the flow can 
become 2D-like when the rotation is sufficiently fast. Recently, \cite{novi2019rapidly} further generalized the situation by varying the tilting angle of the rotation axis with respect to gravity. This configuration mimics the flow at different latitudes in a rotating fluid shell. A large-scale cyclonic vortex tilted along the rotation axis is identified for $\phi$ between $45^\circ$ and $90^\circ$, where $\phi$ is the angle between the rotation axis and the horizontal plane. At moderate latitudes the
calculations of \cite{novi2019rapidly}  suggest the possible coexistence of zonal jets and tilted-vortex solutions.

Even though flows in geophysics, astrophysics, and plasma-physics often occur in large-aspect ratio systems,
most of the previous zonal flow studies with free-slip conditions at the plates
were performed for horizontally periodic 
{\it small}-aspect ratio cells, typically $\Gamma=2$ (2D) or $\Gamma=2\pi$ (3D). However,
 recent studies on {\it large}-aspect ratio 3D RB convection with no-slip plates revealed the existence of superstructures with horizontal extent of six to seven times the height of the domain \citep{hartlep2003large,parodi2004clustering,pandey2018turbulent,stevens2018turbulent,krug2020coherence,green2020resolved}. These findings motivated us to study zonal flow at much larger $\Gamma$ (up to 128) than hitherto done, in order  to test
 whether zonal flow will sustain at these much larger $\Gamma$, or  whether some other large-scale structures evolve, which are not captured in simulations with small $\Gamma$.

We will find that for free-slip plates and periodic boundary conditions the aspect ratio indeed has a very strong influence on the flow phenomena in 2D RB convection and in 3D RB convection with span-wise rotation. In particular, we will show that zonal flow is only stable when the aspect ratio of the system is smaller than a critical value, which depends on $Ra$ and $Pr$; it disappears in large-aspect ratio flow configurations. 

The paper is organized as follows. We first describe the simulation details in Section \ref{sec2}. The 2D results are presented and analysed in Section \ref{sec3}, which is divided into three parts. Section \ref{sec31} demonstrates the disappearance of zonal flow with increasing the aspect ratio $\Gamma$. Section \ref{sec32} studies the coexistence of multiple convection roll states. The effective scaling relations for $Nu(Ra,Pr)$ and $Re(Ra,Pr)$ for different convection roll states are discussed in Section \ref{sec33}. The 3D RB convection with increasing rotation strength about an axis parallel to the plate (i.e., increasing two-dimensionalization according to Taylor-Proudman theorem) is discussed in  Section \ref{sec4}, where we also show the transition from zonal flow to convection roll states with increasing $\Gamma$. We summarize our findings in Section \ref{sec5}.

\begin{table}
\tabcolsep 10pt
\renewcommand{\arraystretch}{1.}
\label{table1}
\begin{center}
\begin{tabular}{|c|c|c|c|c|c|c|c|c|}
\hline
$Ra$ & $Pr$ & $\Gamma$ & $ N_z^r $ & $N_z^z$ & $N_{BL}^r$ & $N_{BL}^z$ \\
\hline 
$10^7$ & 0.71 & 16 & $256 $ & - & 12 & - \\
$10^7$ & 10 & [1,64] & $256 $ & 256 & 13 & 21 \\
$3\times10^7$ & 10 & [1,64] &$256$ & 256 & 11 & 20  \\
$10^8$ & [1,100] & [1,128] &$384$ & 256,384 & 15 & 20 \\
$3\times10^8$ & 10 & [1,64] & $512$ &256,384 & 17 & 24 \\
$10^9$ & 10 & 2,16 &$768$ & 768 & 20 & 46 \\
\hline
\end{tabular}
\end{center}
\caption{Overview of the run simulations in 2D. The first three columns indicate the $Ra$, $Pr$, and $\Gamma$ range of the simulations. $N_z^r$ and $N_z^z$ indicate the number of grid points in the vertical direction for the simulations with initial conditions of roll states and shear flow, respectively. $N_{BL}^r$ and $N_{BL}^z$ indicate the minimum number of grid points in the thermal boundary layer for convection roll states and zonal flow state, respectively. We note that the number of grid points in the boundary layer is always higher than that given by the recommendation of \cite{shishkina2010boundary} for the no-slip case, which is about $5$ to $9$ for this $Ra$ range, to ensure that the boundary layers are resolved. The number of grid points in the horizontal direction is generally equal or larger than $N_x=N_z\times\Gamma$. For $Ra=10^8$ and $3\times10^8$, $N_z^z=256$ is used  only for the large $\Gamma$ cases (fore example, $Ra=3\times10^8,\Gamma=32$ and 64) where very long simulations are performed, in order to test whether the zonal flow state can stably exist.}\label{tab01}
\end{table}

\section{Simulation details}\label{sec2}

\begin{figure}
 \centering
 \begin{overpic}[width=0.93\textwidth]{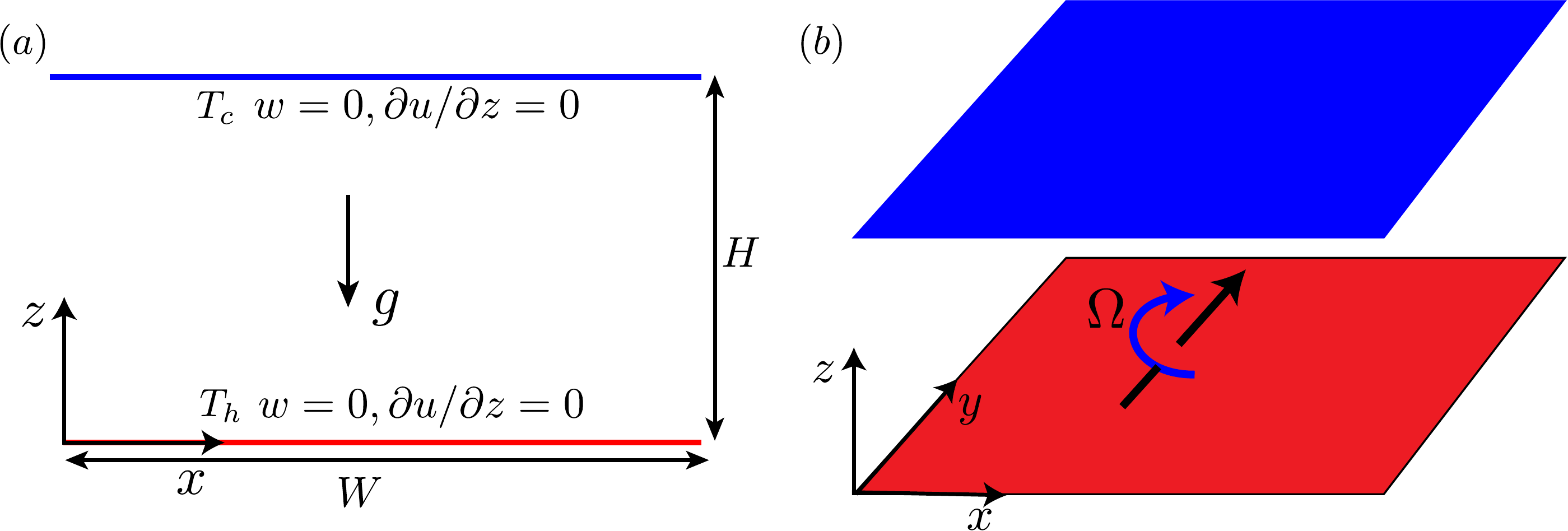}
 \end{overpic}
 \caption{Sketch of (\textit{a}) 2D RB convection and (\textit{b}) 3D RB convection with span-wise rotation for free-slip plates and horizontally periodical conditions.}\label{sketch}
\end{figure}

The configurations and the coordinate systems used in this work are shown in figure \ref{sketch}. We performed direct numerical simulations using the second-order staggered finite difference code AFiD. Details about the numerical method can be found in \citep{verzicco1996finite,van2015pencil,zhu2018afid}. The governing equations in dimensionless form read

\begin{gather}
\nabla\cdot\boldsymbol{u} = 0, \label{eq01} \\
\frac{\partial \boldsymbol{u}}{\partial t} + \boldsymbol{u}\cdot\nabla\boldsymbol{u} = -\nabla p+ \sqrt{\frac{Pr}{Ra}}\nabla^2\boldsymbol{u} -\frac{1}{Ro}{\vec{\boldsymbol{e}}_y}\times \boldsymbol{u} + \theta{\vec{\boldsymbol{e}}_z}, \label{eq02}\\
\frac{\partial \theta}{\partial t} + \boldsymbol{u}\cdot\nabla \theta  = \frac{1}{\sqrt{RaPr}}\nabla^2\theta. \label{eq05}
\end{gather}

\noindent Here $\vec{\boldsymbol{e}}_y$ and $\vec{\boldsymbol{e}}_z$ are the unit vector in the $y$ and $z$ direction, respectively. $u,t,p,\theta$ are velocity, time, pressure and temperature, respectively.  The length and velocity are non-dimensionalized using the height of the convection cell $H$ and the free-fall velocity $U={(g\alpha\Delta H)}^{1/2}$, respectively. This implies as reference time the free-fall time $t_f=H/U$. 
For the 3D simulations also 
 the Rossby number $Ro=U/(2\Omega H)$ is used, where $\Omega$ is the angular velocity. 
   Non-uniform grids with points clustered near the top and bottom plates are employed.

 How to choose the initial conditions to trigger the different flow states?  For the zonal flow simulations we used a linear
 shear-flow velocity profile $u(z)=2z-1,w=0$ in combination with a linear temperature profile $\theta(z)=1-z$ with additional random perturbations as initial conditions. 
 In addition, different convection roll states were
  generated using a Fourier basis: $u(x,z)={\rm sin}(n^{(i)}\pi x/\Gamma){\rm cos}(\pi z), w(x,z)=-{\rm cos}(n^{(i)}\pi x/\Gamma){\rm sin}(\pi z)$ where $n^{(i)}$ indicates the number of initial rolls in the horizontal direction, while the initial temperature is the same as zonal flow simulations. A similar Fourier basis was  also
   used to study heat transport \citep{chong2018effect} and flow reversals \citep{chandra2011dynamics,wang2018flow,wang2019flow,chen2019emergence}. An overview of the 2D simulations and the used grid resolutions are given in table \ref{tab01}. Table \ref{tabra} and \ref{tabpr} in the Appendix list 2D simulation details for the main cases where $Nu$ and $Re$ are discussed.  The 3D simulation details are given in table \ref{tab3d}, where the 2D simulations for the corresponding parameters are also listed for comparison.

\section{Two-dimensional simulations}\label{sec3}
\subsection{Disappearance of zonal flow with increasing $\Gamma$}\label{sec31}

\begin{figure}
 \centering
 \begin{overpic}[width=1\textwidth]{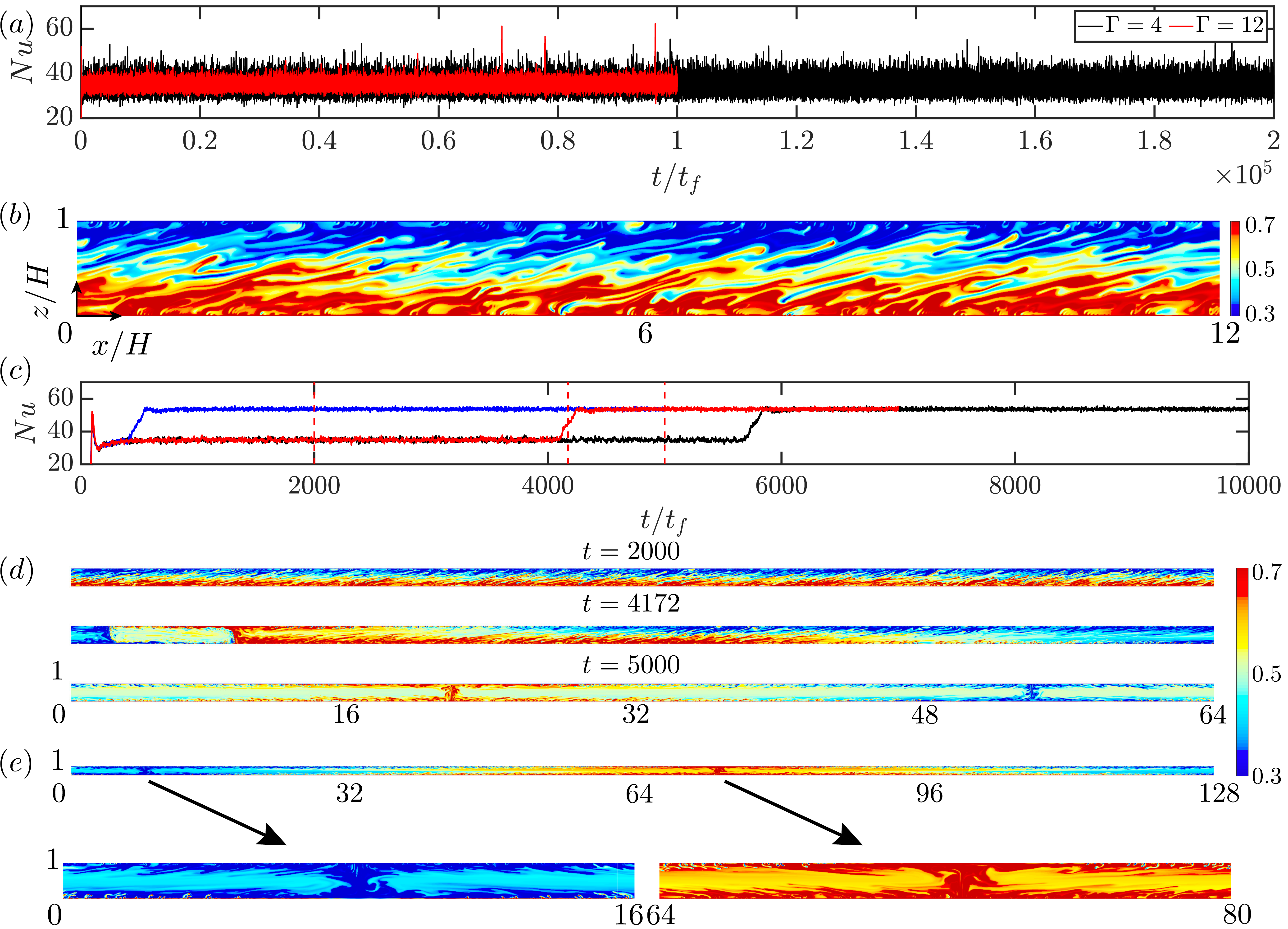}
 \end{overpic}
 \caption{(\textit{a}) Time evolution of $Nu$ for the zonal flow state for $Ra=10^8$ and $Pr=10$ with $\Gamma=4$ (black line) and $\Gamma=12$ (red line). (\textit{b}) Temperature snapshot for the  zonal flow state for $Ra=10^8$, $Pr=10$, $\Gamma=12$. (\textit{c}) Time evolution of $Nu$ for $Ra=10^8$, $Pr=10$, $\Gamma=64$. The three curves correspond to three separate simulations with random perturbations added to the initial temperature field. In all the cases, the flow undergoes a transition from zonal flow to convection roll states, for which $Nu$ is larger. (\textit{d}) Temperature snapshots at different times denoted by 
 the red dashed lines for the simulation indicated by the red curve in panel (\textit{c}). At $t=2000$, there is zonal flow, whereas later it features an increasing number of turbulent convection rolls. (\textit{e}) The final two-roll state for $Ra=10^8$, $Pr=10$, $\Gamma=128$, and the zoom in of the two plume-ejecting regions. For all these simulations the initial velocity had a 
 linear shear flow profile $u(z)=2z-1,w=0$, in order to trigger a zonal flow state.}\label{r8ar64}
\end{figure}

We first show what will happen to zonal flow with increasing $\Gamma$. From \cite{goluskin2014convectively}, 
it is  known  that zonal flow exists  for $Ra=10^8$, $Pr=10$, $\Gamma=2$. With increasing $\Gamma$, we find that 
for this $Ra$ and $Pr$, zonal flow can stably exist for $\Gamma\le12$. In figure \ref{r8ar64}($a$) we show that zonal flow is statistically stable at least up to $200\ 000$ free-fall time units for $\Gamma=4$ and at least up to $100\ 000$ free-fall time units for $\Gamma=12$. Here, we used ``statistically stable''  to denote that the corresponding chaotic flow state is always sustained in our long-time simulations. Temperature snapshot of the zonal flow for $\Gamma=12$ in figure \ref{r8ar64}($b$) demonstrates that hot plumes drift leftwards, and 
 cold plumes drift rightwards. This produces a strong horizontal shear in which however the vertical heat transport is low.

\begin{figure}
 \centering
 \begin{overpic}[width=0.95\textwidth]{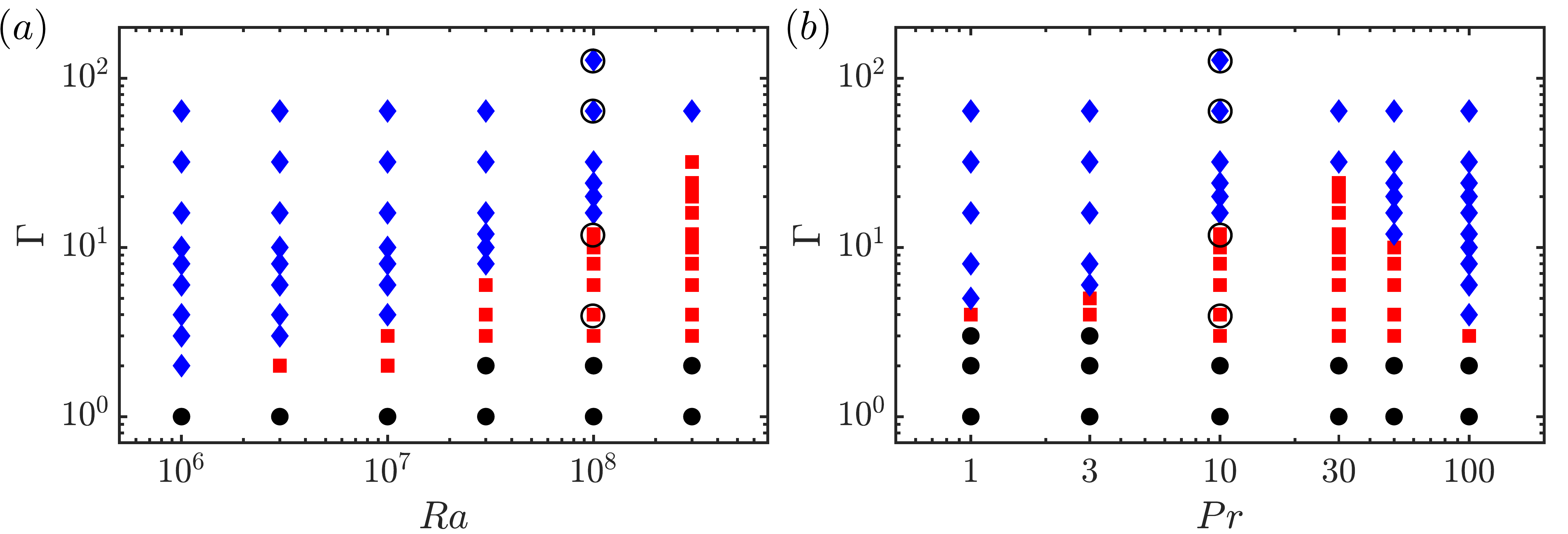}
 \end{overpic}
 \caption{Phase diagram in the (\textit{a}) $Ra-\Gamma$ parameter space for $Pr=10$ and  in the (\textit{b}) $Pr-\Gamma$ parameter space for $Ra=10^8$. Black circles (\large$\bullet$) correspond to only zonal flow, red squares (\textcolor{red}{$\blacksquare$}) denote coexistence of zonal flow and convection rolls, and blue diamonds (\textcolor{blue}{$\blacklozenge$}) indicate that only convection roll states are stable. The black hollow circles mark the cases shown in figure \ref{r8ar64}.}\label{phase}
\end{figure}

We now explore even larger aspect ratio domains. Figure \ref{r8ar64}($c$) shows the time evolution of $Nu$ for three separate simulations with random perturbations added to the initial temperature field for $Ra=10^8$ and $Pr=10$ in a $\Gamma=64$ cell. For all the three simulations, the zonal flow eventually evolves to a convection roll state. The time at which the transition occurs is very different for each simulation. The reason is that the flow is susceptible to small differences in the initial conditions, which are different for each simulation due to the random perturbations to the initial temperature field. Such sensitivity to the initial conditions is typical for chaotic systems and makes it impossible to predict when the transition happens.

Figure \ref{r8ar64}($d$) displays three temperature snapshots at different time instants indicated by the red dashed lines in figure \ref{r8ar64}($c$). A complementary movie, showing how the zonal flow undergoes a  transition towards a convection roll state, is given in the supplementary material. Initially, the hot plumes travel leftwards and the cold plumes to the right. The transition starts when some local hot plumes are strong enough to deviate upwards and cross the whole fluid layer up to the collision with the upper cold plate. This prevents the further rightward motion of the neighbouring cold plumes, which instead start to move downwards. This process generates a local large-scale circulation, as is observed in the temperature field at $t=4172$. The circulation grows over time until two stable convection rolls of equal size are formed as seen in the temperature filed at $t=5000$. Figure \ref{r8ar64}($e$) shows that we also obtain a two-roll state for $\Gamma=128$, such that the horizontal extent of the convection roll is $64$ times the height of the convection cell, and this state is stable for more than $10\ 000$ free-fall time units.

We now explore the phase diagram for the different flow states in the parameter space spanned by $Ra$, $Pr$, and $\Gamma$, see figure \ref{phase} for the simulated cases. We find that in small-aspect ratio cells, only zonal flow is stable, while in large-aspect ratio cells, only convection roll states are stable. For intermediate aspect ratios, we find a regime in which both zonal flow and convection roll states are stable, depending on the initial conditions mentioned in Section \ref{sec2}. We call this regime the bi-stable one. In order to map out an accurate phase diagram, we performed
 long simulations for the bi-stable cases with largest $\Gamma$ to {\color{blue}{conclude}}
 that the corresponding zonal flow can stably exist and does not evolve to a convection roll state. Overall, we 
 ran the simulations at least $50\ 000$ free-fall time units for these cases, which corresponds to at least 5 viscous diffusive time units ($H^2/\nu$) or 0.5 thermal diffusive time units ($H^2/\kappa$). 
 From figure \ref{phase}($a$) it can be seen that, when $Ra$ is increased, 
  the bi-stable regime 
 exists in an increasing $\Gamma$ range. This is also consistent with the finding that zonal flow develops more readily for higher $Ra$ for $\Gamma=2$ \citep{goluskin2014convectively}. Figure \ref{phase}($b$) demonstrates that the $\Gamma$ range for the bi-stable state also depends on $Pr$. 
 For $Ra = 10^8$ 
 the largest range of bi-stable state exists at $Pr\sim30$, namely between $\Gamma=3$ and $\Gamma=24$. 

We have already shown that zonal flow can not be sustained, and only convection roll states are observed, when $\Gamma$ is larger than a critical value, which depends on $Ra$ and $Pr$. A related question is how many convection rolls (in other words, what is the mean aspect ratio of  individual convection roll) can develop for a specific ($Ra$, $Pr$, $\Gamma$). In the next subsection, we will explore the possible convection roll states using different initial roll states generated by different
Fourier basis,  as explained  in Section \ref{sec2}.

\subsection{Coexistence of multiple convection roll states at large $\Gamma$} \label{sec32}

\begin{figure}
 \centering
 \begin{overpic}[width=0.75\textwidth]{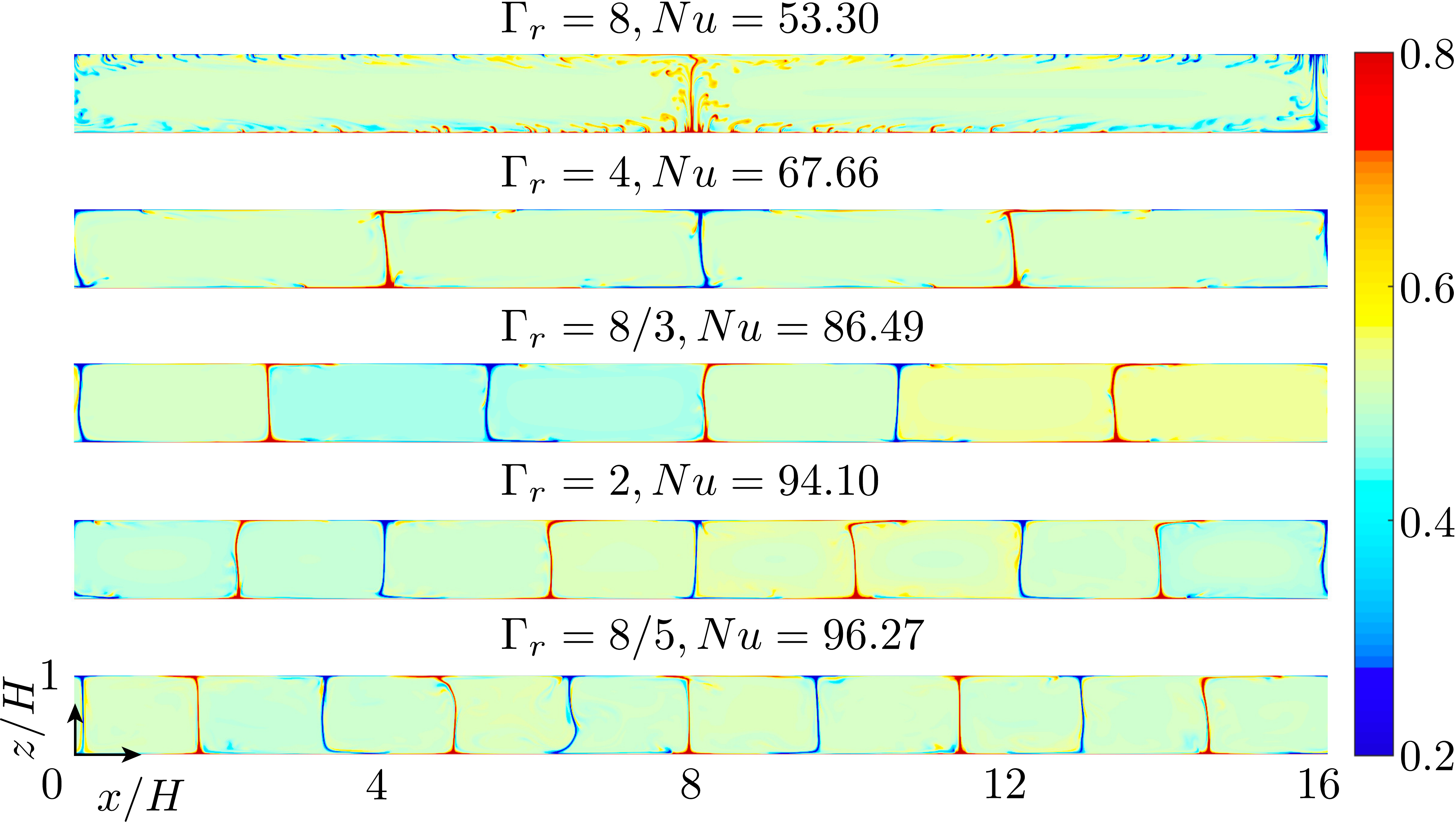}
 \end{overpic}
 \caption{Temperature snapshots of different roll states for $Ra=10^8$ and $Pr=10$ in a $\Gamma=16$ periodic cell. The dimensionless mean horizontal size of the convection roll $\Gamma_r$ (i.e., the mean aspect ratio of one individual roll) and the Nusselt number $Nu$ for each state are indicated. The different roll states are from initial conditions with different number of initial rolls. All these states can stably exist for long time (see table \ref{tabra} in Appendix) without undergoing a transition to other states.}\label{ar16roll}
\end{figure}

In this subsection, we study the coexistence of multiple convection roll states in large-aspect ratio domains, all being statistically stable states once achieved. Figure \ref{ar16roll} shows that for $Ra=10^8$, $Pr=10$, and free-slip at the plates, 
in a $\Gamma=16$ system convection rolls with a mean dimensionless horizontal size of $1.6\leq\Gamma_r\leq8$ are all 
statistically stable. The heat transport considerably increases 
 with decreasing mean aspect ratio $\Gamma_r$ of an individual convection roll.  
 For example, the heat transport for the $\Gamma_r=1.6$ state is almost twice as high as that for $\Gamma_r=8$. 
 Although  
 it had been observed before that convection rolls with smaller $\Gamma_r$ imply 
  a higher heat transport -- e.g. 
  for 2D RB convection with no-slip plates \citep{van2012flow,wang2018multiple}, for RB convection in an annulus convection cell \citep{xie2018flow}, and also for  Taylor-Couette flow \citep{huisman2014multiple} -- 
  in those cases 
the observed increase in the transport is typically a couple of percents, and by far not 
 as large as observed for RB with free-slip plates and large aspect ratio cells as studied here. This difference is due to different plume dynamics and the associated spatial dependence of the local Nusselt number,   $Nu(x)$ , as  we will  discuss
  later.

 \begin{figure}
 \centering
 \begin{overpic}[width=1\textwidth]{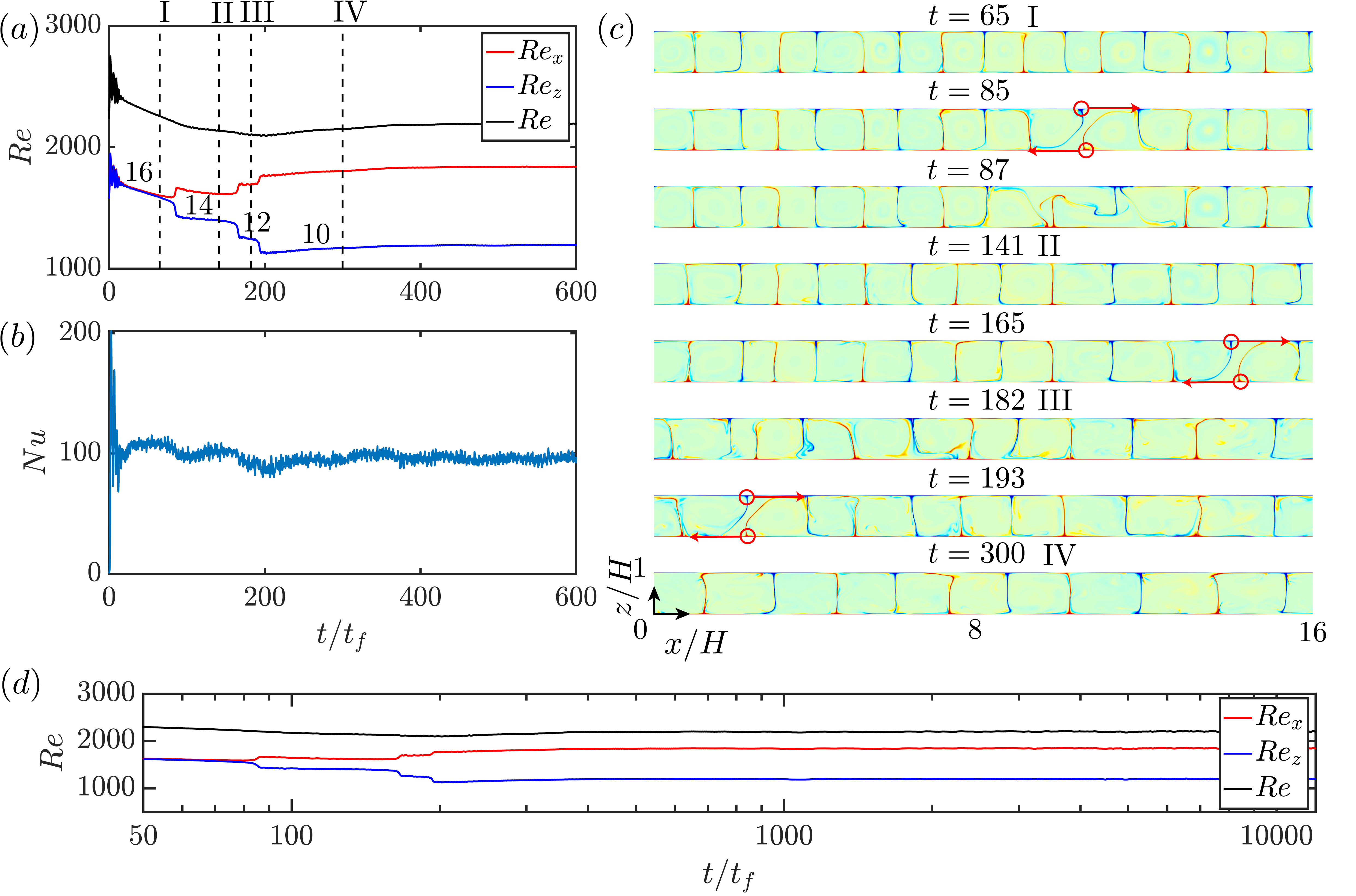}
 \end{overpic}
 \caption{Time evolution of (\textit{a}) $Re$ and (\textit{b}) $Nu$ for $Ra=10^8, Pr=10, \Gamma=16$ with an initial  sixteen-roll state. $Re_x=\sqrt{(Ra/Pr)}\sqrt{(\left<u^2\right>_V)}$ is horizontal Reynolds number and $Re_z=\sqrt{(Ra/Pr)}\sqrt{(\left<w^2\right>_V)}$ the vertical one. (\textit{c}) Temperature snapshots at different times. The roll merging can be seen, namely the flow undergoes a transition from initial sixteen-roll state (\uppercase\expandafter{\romannumeral1}), to a fourteen-roll state (\uppercase\expandafter{\romannumeral2}), to a twelve-roll state (\uppercase\expandafter{\romannumeral3}), and then to the final ten-roll state (\uppercase\expandafter{\romannumeral4}). The figure has the same colour scale as figure \ref{ar16roll}. (\textit{d}) Time evolution of $Re$ for much longer time (on a log-scale) to show that the final ten-roll state is stable without undergoing a transition to an other roll state. }\label{roll_merge}
\end{figure}

\begin{figure}
 \centering
 \begin{overpic}[width=0.95\textwidth]{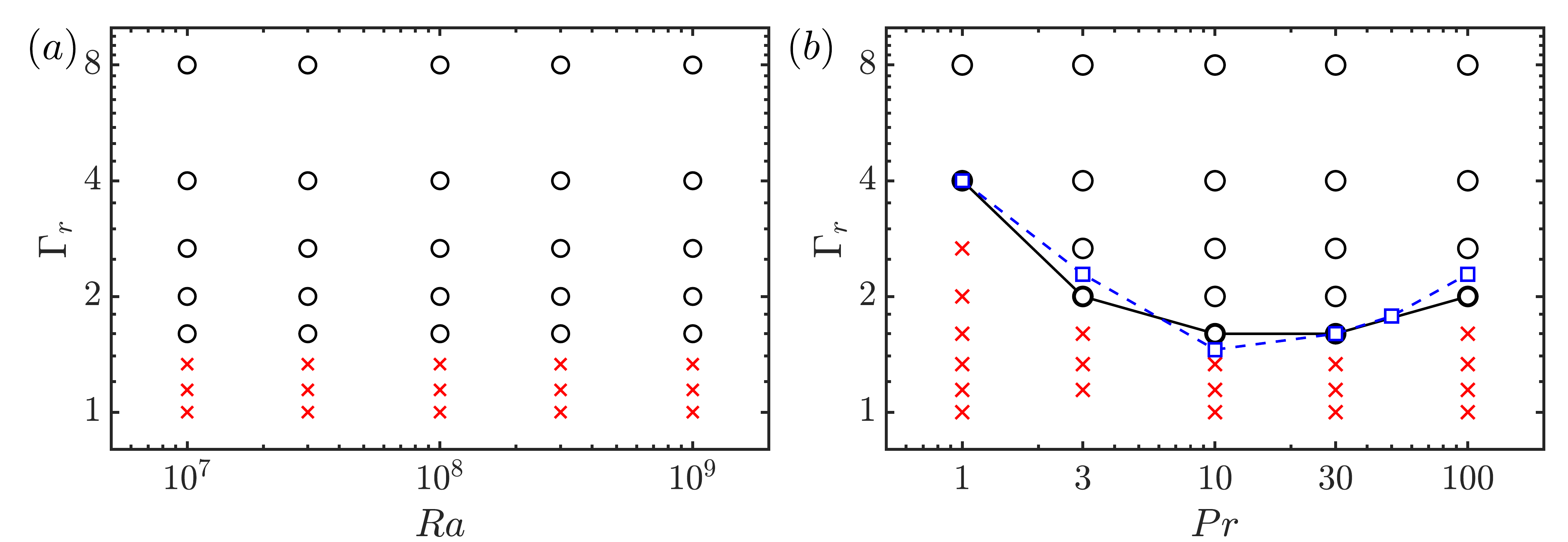}
 \end{overpic}
 \caption{Phase diagram for different roll states for (\textit{a}) $Pr=10$, $\Gamma=16$ and (\textit{b}) $Ra=10^8,\Gamma=16$. Circles denote that the corresponding roll state with the mean aspect ratio $\Gamma_r$ of an individual roll is stable, while crosses denote that the roll state is not stable. The solid line in (\textit{b}) connects the minimal mean aspect ratio $\Gamma_{r,\rm{min}}$ of an individual convection roll for different $Pr$ for $\Gamma=16$, while the dashed line connects $\Gamma_{r,min}$ for different $Pr$ for $\Gamma=32$.}\label{pr10_roll}
\end{figure}

We also tested 
 initial conditions with 12, 14, 16 rolls for $Ra=10^8,Pr=10,\Gamma=16$.
  These states with smaller rolls are not stable and will finally undergo a transition to the ten-roll state by merging of convection rolls.
  From figure \ref{roll_merge}($a$) is is seen that the vertical Reynolds number $Re_z$ has a sudden decrease during merging of rolls,  because the strong vertical motion is concentrated near the plume-ejecting regions between two neighbouring rolls.
   The decrease of $Nu$ during merging events can also be observed in figure \ref{roll_merge}($b$), which is related to the decreased vertical motion. Figure \ref{roll_merge}($c$) shows how the flow undergoes a transition from the initial sixteen-roll state to the final ten-roll state by successive merging of 
   convection rolls. The transition happens when the balance of the roll state is broken by horizontal motion of local hot/cold plumes: 
  In the second  snapshot  at $t=85$ the system is still in the initial sixteen role state. However,
  one can already see that a local hot plume moves leftwards while its neighbouring cold plume moves rightwards (marked by
  red arrows). In the third snapshot at $t=87$
   two hot  plumes merge to  a single one and so do two cold ones, thus annihilating two rolls.
    The resulting fourteen-roll state is  shown in the fourth snapshot taken at $t=141$. 
    At later times the horizontal motion of the plumes and their further merging let  the fourteen-roll state  evolve 
     to a twelve-roll state ($t=182$,  \uppercase\expandafter{\romannumeral3}), and finally to a ten-roll state ($t=300$,  \uppercase\expandafter{\romannumeral4}). We also performed very 
      long simulations as indicated in figure \ref{roll_merge}($d$),  from which we conclude 
       that the ten-roll state can statistically stably exist for a very long time without undergoing any further  transition to 
      yet  another state.

Figure \ref{pr10_roll} displays phase diagrams for all the possible convection roll states in the 
$Ra-\Gamma_r$ and $Pr-\Gamma_r$ parameter spaces. The stable roll states can last for several thousand free-fall time units without undergoing a transition to other states (see Appendix). Figure \ref{pr10_roll}($a$) shows a weak dependence of $\Gamma_r$ on $Ra$. One can observe the same stable roll states for the considered $Ra$ range. In contrast, a pronounced dependence of $\Gamma_r$ on $Pr$ is observed in figure \ref{pr10_roll}($b$), where 
 convection roll states with the smallest $\Gamma_r$ are observed for intermediate $Pr\approx10$. The minimum  $\Gamma_r=16/11$ occurs 
  for $\Gamma=32$, which means that the horizontal extent of a stable convection roll is always larger than the height of the system. This explains why convection rolls  cannot be supported for small $\Gamma\approx2$,  where indeed
 only  zonal flow was obtained. From figure \ref{pr10_roll}($b$) it can also be concluded
   that these results are independent of the aspect ratio of the system once it is large enough, as we obtained
    almost the same result for $\Gamma=16$ and $\Gamma=32$ domains.

\begin{figure}
 \centering
 \begin{overpic}[width=0.96\textwidth]{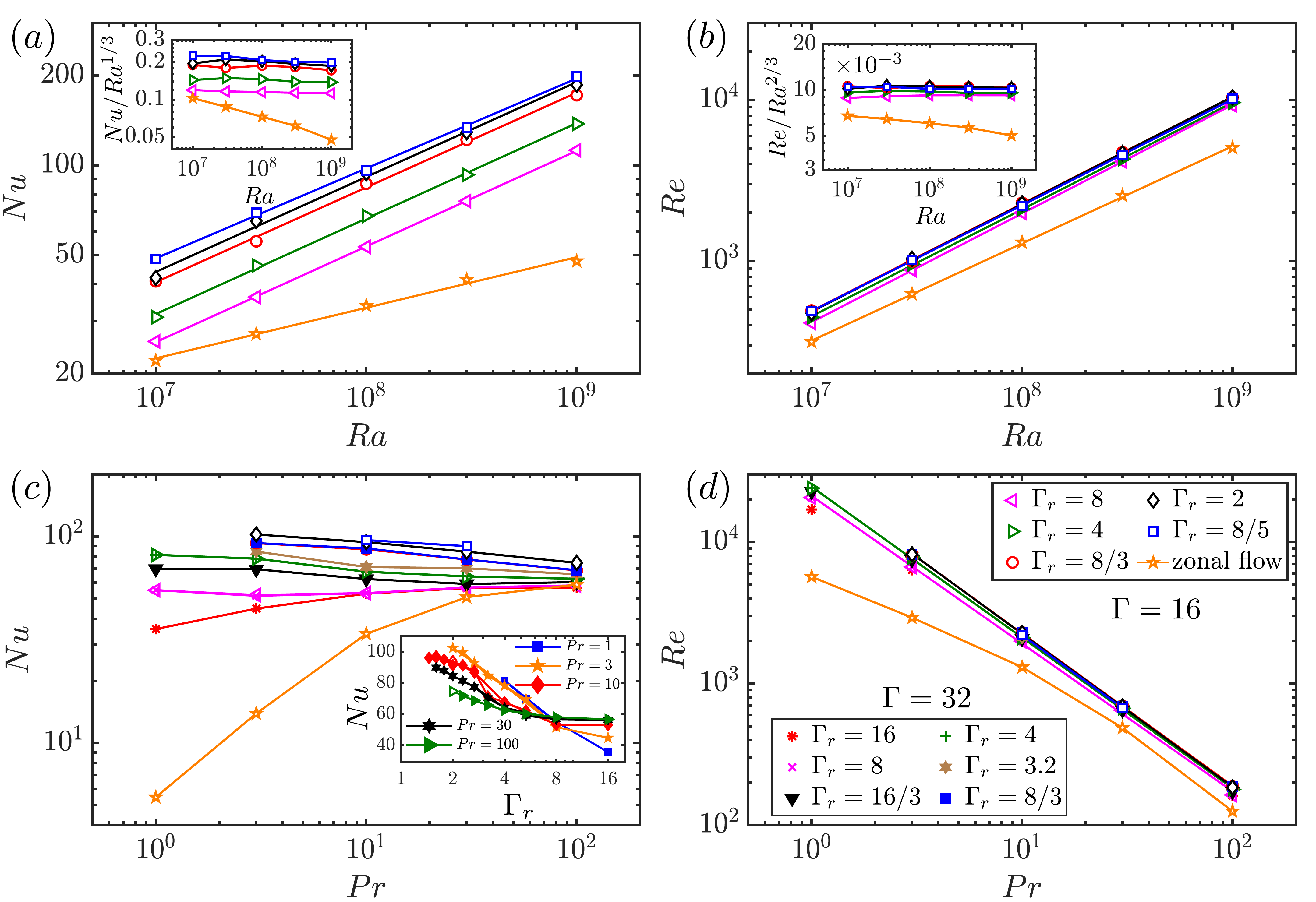}
 \end{overpic}
 \caption{ (\textit{a}) $Nu$ and (\textit{b}) $Re$ as function of $Ra$ for different convection roll states 
 (see legend in panel (d)) 
 for $Pr=10$, $\Gamma=16$ and the zonal flow state (see orange stars on solid orange line) 
 for $Pr=10$, $\Gamma=2$. (\textit{c}) $Nu$ and (\textit{d}) $Re$ as function of $Pr$ for different roll states for $Ra=10^8$
 (for $\Gamma=16$  and $\Gamma = 32$) 
  and for the zonal flow state, for which we put  $\Gamma=2$. 
 Again, see the legend in panel (d). 
 The inset in  (\textit{c}) shows $Nu$ as function of $\Gamma_r$ for $Ra=10^8$ and different $Pr$ obtained for $\Gamma=16$ (hollow symbols) and $\Gamma=32$ (solid symbols). The solid symbols often overshadow the hollow ones as the symbol sizes are the same. }
 \label{scaling}
\end{figure}

\subsection{Nusselt number and Reynolds number} \label{sec33}

We now discuss the effective scaling relations of $Nu$ and $Re$ as function of $Ra$ and $Pr$. These relationships are usually expressed with effective scaling laws $Nu\sim Ra^{\gamma_{Nu}}Pr^{\alpha_{Nu}}$ and $Re\sim Ra^{\gamma_{Re}}Pr^{\alpha_{Re}}$ \citep{ahlers2009heat}. The effective scaling laws have been widely discussed for no-slip cases for both 2D and 3D convection \citep{ahlers2009heat}. For the 2D horizontally periodic cases with no-slip plates, $Nu\sim Ra^{0.29}$ is found with $Pr=1,Ra\le10^{10}$ \citep{johnston2009comparison,zhu2018transition}. For 2D RB convection with no-slip plates and sidewalls with unit aspect ratio, several studies have shown that $Nu\sim Ra^{0.3}$ and $Re\sim Ra^{0.6}$ \citep{sugiyama2009flow,zhang2017statistics,wang2019penetrative}.  However, how these effective scaling relations
 will change for free-slip plates has not been explored, especially not for convection roll states, which are only present in large enough domain size.

\begin{table}
\begin{center}
\tabcolsep 17pt
\renewcommand{\arraystretch}{1.}
\begin{tabular}{|c|c|c|c|c|c|c|c|c|}
\hline
Flow state  & $\gamma_{Nu}$  & $\gamma_{Re}$     & $\alpha_{Re}$\\
\hline
$\Gamma_r=8$  & 0.321 & 0.675      & -1.043 \\
$\Gamma_r=4$ & 0.320 & 0.663     & -1.065 \\
$\Gamma_r=8/3$ & 0.318 & 0.665   &-1.077\\
$\Gamma_r=2$  & 0.318 & 0.667    &-1.082\\
$\Gamma_r=1.6$ & 0.302 & 0.657 & -1.078\\
Zonal flow & 0.170  & 0.603     & -\\
\hline
\end{tabular}
\caption{The effective scaling exponents for fitted effective scaling relations
 $Nu \sim Ra^{\gamma_{Nu}}$, $Re \sim Ra^{\gamma_{Re}}$ and $Re \sim Pr^{\alpha_{Re}}$ for different roll states for $Pr=10$ obtained in an aspect ratio $\Gamma=16$ domain. The zonal flow data is for $\Gamma=2$.}\label{tbscaling}
\end{center}
\end{table}

Figure \ref{scaling}($a$) and \ref{scaling}($b$) show $Nu$ and $Re$ as function of $Ra$ for both zonal flow ($\Gamma=2$) and convection roll states ($\Gamma=16$) for $Pr=10$. Figure \ref{scaling}($a$) reveals that the heat transfer in the convection roll state is much higher than that of the zonal flow state. Detailed information about the obtained scaling exponents is listed in table \ref{tbscaling}. For the convection roll states we find that the effective 
scaling exponent $\gamma_{Nu}$ in $Nu \sim Ra^{\gamma_{Nu}}$ is approximately $0.3$.
It  increases with increasing $\Gamma_r$, reaching about 1/3 for the largest $\Gamma_r=8$, which is the value predicated by the Grossmann-Lohse (GL) theory 
for the $\rm{\uppercase\expandafter{\romannumeral1}}_\infty^<$ and $\rm{\uppercase\expandafter{\romannumeral3}}_\infty$ regimes \citep{gl2000,grossmann2001thermal,shishkina2017scaling} for the
no-slip case. For zonal flow $\gamma_{Nu}$ is much smaller, namely only  $0.17$. The effective
scaling exponent $\gamma_{Re}$ in $Re \sim Ra^{\gamma_{Re}}$ is about $0.6$ for zonal flow and  about $0.67$ for the convection roll state, which is also close to the GL predication of 2/3 for the $\rm{\uppercase\expandafter{\romannumeral1}}_\infty^<$ and $\rm{\uppercase\expandafter{\romannumeral3}}_\infty$ regimes \citep{grossmann2001thermal,shishkina2017scaling} for no-slip cases, while it is larger than $0.6$ that has been reported for 2D RB convection with no-slip plates \citep{sugiyama2009flow,zhang2017statistics,wang2019penetrative}.

 Next, we will discuss the Prandtl number dependence of the Nusselt number, $Nu(Pr)$. 
 For no-slip plates in 3D cases, it is known that the $Nu$ is maximal around $Pr\sim2-3$, and after that it declines with increasing $Pr$ \citep{ahlers2001prandtl,xia2002heat,stevens2011prandtl}. This remarkable maximum in $Nu(Pr)$  had
  in fact been  predicted before by the GL theory \citep{gl2000,grossmann2001thermal}.
  In contrast, 
 for the 2D cases, \cite{huang2013counter} showed  that $Nu(Pr)$ has a minimum, rather than maximum as in the 3D case, 
 namely at $Pr\sim2-3$ for moderate $Ra$. This anomalous relation is caused by counter-gradient heat transport in 2D cases. 
 
 How does the $Nu(Pr)$ dependence look like for the 2D RB case with 
 free-slip plates? For the zonal flow state
  \cite{goluskin2014convectively} already showed that $Nu$ is an increasing function of $Pr$ in the range $1\le Pr\le10$. 
  Figure \ref{scaling}($c$) and \ref{scaling}($d$) show the 
  relations for $Nu(Pr)$ and $Re(Pr)$ for all states with free-slip plates (i.e., both for zonal flow and for various convection roll states), 
  where we included both $\Gamma=16$ and 32 data for the convection roll state.  
 From  figure \ref{scaling}($c$) it can  be seen 
 that the $Nu(Pr)$  trend shown by \cite{goluskin2014convectively} 
 is also valid for the wider range of $Pr$ analysed in this paper. The reason why $Nu$ is much smaller for small $Pr$ is that zonal flow features  intermittent bursts whereas most of the time  $Nu$ is around 1 \citep{goluskin2014convectively}. For large $Pr$, the flow does not burst and convective heat transport with $Nu\gg1$ is sustained at all times, thus the corresponding $Nu$ is larger than that of the small $Pr$ cases. 
 
 For the convection roll states, figures \ref{scaling}($c$) and \ref{scaling}($d$) reveal that both $Nu$ and $Re$ collapse well for the same mean aspect ratio of an individual convection cell ($\Gamma_r=8, 4, 8/3$) for both $\Gamma=16$ and 32. 
 This suggests
  that the global transport properties are directly related to the mean convection roll size $\Gamma_r$. Figure \ref{scaling}($c$) also shows that $Nu$ increases monotonically with increasing $Pr$ for large mean convection roll size $\Gamma_r=16$. This can be interpreted as that the flow for the large $\Gamma_r=16$ cases can be viewed as ``localised'' zonal flow, thus the $Nu(Pr)$ follows the similar trend as that of zonal flow. In contrast, for small $\Gamma_r$,
   $Nu$ decreases with increasing $Pr$. The different $Nu(Pr)$ trends for large and small $\Gamma_r$ can also be seen in the inset of  figure \ref{scaling}($c$). For $Re(Pr)$, figure \ref{scaling}($d$) shows that the $Re$ follows $Re\sim Pr^{-1}$ for convection roll states (see table \ref{tbscaling}), the exponent $-1$ is the same as that of the GL predication for the $\rm{\uppercase\expandafter{\romannumeral1}}_\infty^<$ and $\rm{\uppercase\expandafter{\romannumeral3}}_\infty$ regimes for no-slip cases.

\begin{figure}
 \centering
 \begin{overpic}[width=0.93\textwidth]{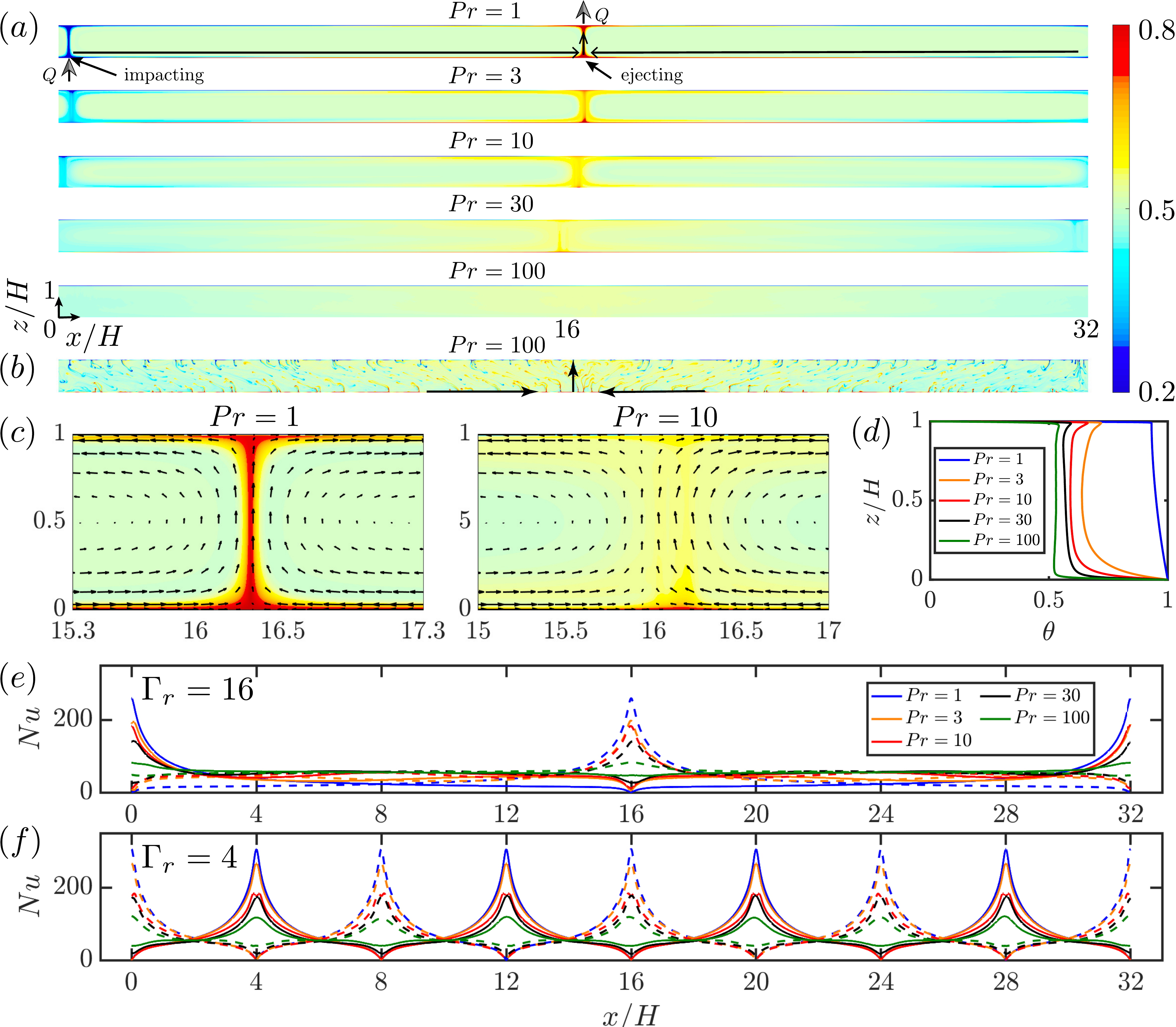}
 \end{overpic}
 \caption{ (\textit{a}) Time-averaged temperature fields for the $\Gamma_r=16$ roll state for different $Pr$ at $Ra=10^8$ obtained in a $\Gamma=32$ domain. (\textit{b}) Instantaneous temperature field for $Pr=100$ for the $\Gamma_r=16$ state with $Ra=10^8,\Gamma=32$. (\textit{c}) Zoom-in for the time-averaged temperature fields in (\textit{a}) for $Pr=1$ and 10.   (\textit{d}) Temperature profiles for the different $Pr$ at the centre point of plume-ejecting regions ($x/H\approx16$) where the
 local $Nu$ is smallest.  Panels (\textit{e}) and (\textit{f})  show the spatial dependence of $Nu(x)$  at the hot plate at $z=0$ (solid lines) and the cold plate at $z=1$ (dashed  lines) for different $Pr$ for the (\textit{e}) $\Gamma_r=16$ and the (\textit{f}) $\Gamma_r=4$ roll states. Note that all curves are shifted such that the minimum local $Nu$ at the hot plate is located at $x/H=16$. }\label{nu_local}
\end{figure}

In order to understand different $Nu(Pr)$ dependence for large and small $\Gamma_r$ as shown in figure \ref{scaling}($c$), we first look at the flow organizations for convection roll states for different $Pr$. Figure \ref{nu_local}($a$) gives the time-averaged temperature fields for the $\Gamma_r=16$ roll state for different $Pr$. The flow near the bottom plate can be divided into plume-ejecting region, plume-impacting region, and between them there is wind-shearing region which occupies large fraction of the domain. In the ejecting region, thermal plumes are emitted, while in the impacting region, the boundary layer is impinged by the plumes from the opposite plate. The wind-shearing region is sheared by the
large-scale circulation. The impacting regions on the top plate are the opposite of the ejecting regions of the bottom plate and vice versa. This kind of division is also adopted in periodic 2D RB convection with no-slip plates \citep{van2015logarithmic}. A remarkable observation is the stable-stratification near the plume impacting region. Figure \ref{nu_local}($c$) shows a
zoom-in of the regions where hot plumes are ejected for $Pr=1$ and 10. When hot fluid impinges the cold plate, it does not have sufficient time to cool down, before  it moves horizontally.
 The consequence is that the temperature of the fluid 
 between  the top boundary layer and the bulk  is  higher than that of the bulk fluid,
  thus implying a 
   stable-stratification. This behavior is even observed at the centre line of the hot plume (the vertical line at the horizontal location where the local bottom wall heat flux is minimal) as is shown in temperature profiles in  figure \ref{nu_local}($d$), 
   where   for $Pr>1$ stable-stratification near the cold plate is observed. The stable-stratification has also been observed at the axis in cylindrical RB convection \citep{tilgner1993temperature,brown2007temperature,wan2019effect}, and  in 2D RB convection with no-slip plates and sidewalls for unit aspect ratio in the central region near the plates \citep{wan2020non}.  The instantaneous temperature fields shown in figure \ref{nu_local}($b$) for $Pr=100$ reveals the ``localised'' zonal flow structures. It can be seen 
    that plumes are ejected everywhere while they can only move vertically and impinge the cold plate in the central region.

Next, we focus on the local properties of the wall heat flux.  The local  wall heat flux is 
expressed through the  local wall Nusselt number $Nu(x)|_{z=0,1}=\partial \left<\theta\right>_t/\partial z|_{z=0,1}$. Figure \ref{nu_local}($e$) and \ref{nu_local}($f$) show the spatial $Nu(x)$ dependence at the plates 
 for $Ra=10^8$ and $ \Gamma=32$ for $\Gamma_r=16$ and $\Gamma_r=4$, respectively. For small $Pr=1$, one sees from figure \ref{nu_local}($a$) the accumulation of hot fluid in the ejecting region near the bottom plate, which cause small temperature gradient (see figure \ref{nu_local}($d$)), and correspondingly small local $Nu$ (see figure \ref{nu_local}($e$)). So the centre of the ejecting region can be denoted as the point where local wall heat flux is minimal. In contrast, for the impacting region ($x/H\approx0$ at the bottom plate) where cold fluid directly impinges the hot plate, there is a
 sharp temperature gradient and thus large local $Nu$ (see figure \ref{nu_local}($e$)). Similar $Nu$ behavior is also observed in the ejecting/impacting regions near the top plate (dashed lines in figure \ref{nu_local}($e$)). The
  physical interpretation is as follows: The heat is ejected into the system through the bottom plate mainly at the plume-impacting regions where local temperature gradient is large, and then it is advected by large-scale circulation to the plume-ejecting regions, where conductive heat flux is low on the wall while convective heat flux is high above the wall. The heat is mainly removed from the system when hot plume impinges the cold plate.

For $\Gamma_r=16$, there is only one impacting region near the bottom plate, and the heat input is still dominated by the wind-shearing region, which occupies a large fraction of the domain. As the wind-shearing region is like ``localised" zonal flow  where $Nu$ increases with increasing $Pr$, the global $Nu$ thus also increases with increasing $Pr$. In contrast, for smaller $\Gamma_r$, there are more impacting regions on the bottom plate, and these impacting regions contribute significantly to the global heat input. As the heat flux at the impacting region increases with decreasing $Pr$, we thus see that the global $Nu$ also increases with decreasing $Pr$.

\section{Three-dimensional simulations} \label{sec4}

We have already shown that the zonal flow observed in 2D RB convection with free-slip plates and horizontally periodic boundary conditions for $\Gamma=2$ \citep{van2014effect,goluskin2014convectively} eventually disappears with increasing $\Gamma$.
What about in 3D, under the same conditions?
 For the 3D RB convection with free-slip plates, previous studies have not reported zonal flow \citep{petschel2013dissipation,kunnen2016transition}. However,
  if we introduce {\it span-wise rotation}  as illustrated in figure \ref{sketch}($b$) where the rotating axis is parallel to the $y$ axis, the flow will  become 2D-like at sufficiently fast rotation, due to the Taylor-Proudman theorem. In this way we may observe zonal flow at certain parameters,  as indeed 
  was  already reported in \citep{von2015generation}. Therefore here we will
  study  span-wise rotating RB convection, focusing on 
  the transition from zonal flow to convection roll states 
   with increasing  aspect ratio $\Gamma$ of the container.

\begin{figure}
 \centering
 \begin{overpic}[width=1\textwidth]{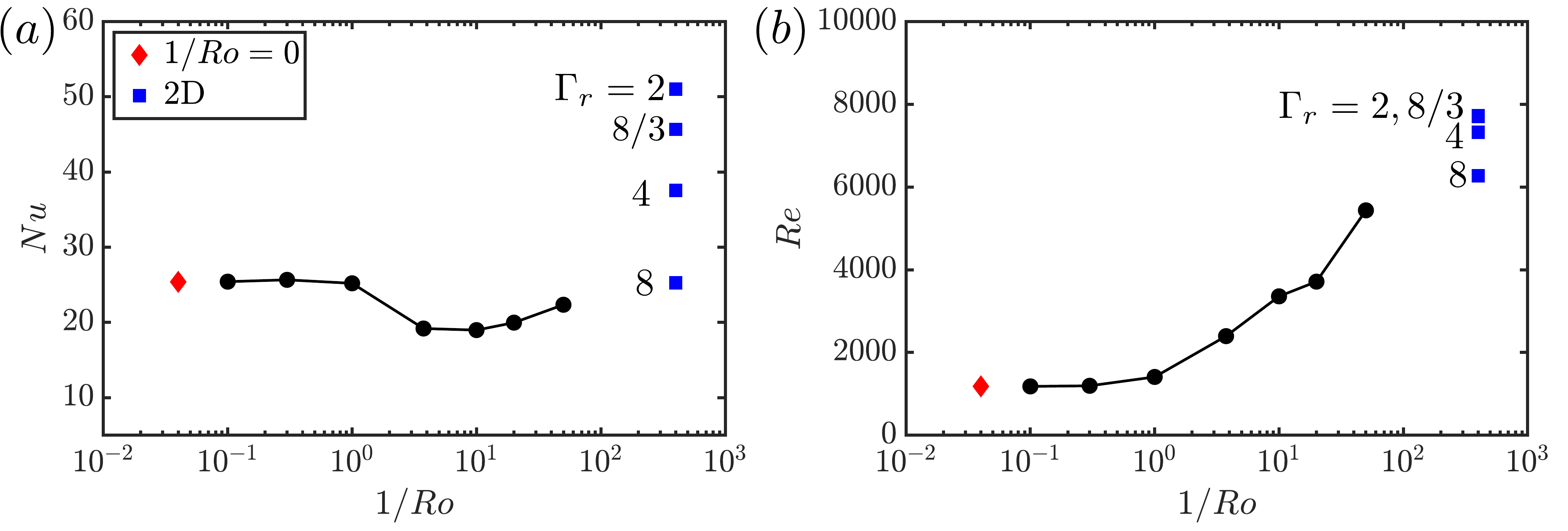}
 \end{overpic} 
 \caption{3D RB convection with span-wise rotation: (\textit{a}) $Nu$ and (\textit{b}) $Re$ as function of  $1/Ro$ for $Ra=10^7,Pr=0.71,\Gamma=16$ (black circles). 
 For orientation with respect to the Nusselt number, the data for non-rotation ($1/Ro=0$, red diamond) and the 2D cases with the same control parameters ($Ra=10^7$, $Pr=0.71$, $\Gamma=16$) for different roll aspect ratios $\Gamma_r$ (blue squares) are also shown; for these data points the value at the $1/Ro$ axis
has no meaning. The Reynolds number $Re$ for the $\Gamma_r=8/3$ (7702.85) and $\Gamma_r=2$ (7726.20) states are close to each other and can not be differentiated in the figure.}\label{3d_nu}
\end{figure}

We first show that both the global transport properties like the Nusselt number $Nu$ and the Reynolds number $Re$, as well as the flow organization increasingly behave like 2D cases with increasing rotation rate. We fix the Rayleigh number to $Ra=10^7$ and the Prandtl number to $Pr=0.71$. To be on the
 safe side, we choose a large domain with $\Gamma=16$, as previous studies showed that large aspect ratios are needed in order to capture the superstructures which have a horizontal size of 6-7 times the height of the domain for 3D RB convection with no-slip plates \citep{pandey2018turbulent,stevens2018turbulent,krug2020coherence,green2020resolved}. The initial conditions have zero velocity and a linear temperature profile for these simulations. For RB convection rotating about a vertical axis for small $Pr=0.71$ with no-slip plates, $Nu$ initially does not change much with increasing the rotation rate (denoted by the inverse Rossby number $1/Ro$),  until after $1/Ro\ge1$,  $Nu$ drops monotonically with increasing $1/Ro$ \citep{zhong2009prandtl}. Figure \ref{3d_nu}($a$) shows that for span-wise rotating RB convection $Nu$ also initially does not change much for $1/Ro\le1$.  After that $Nu$ drops monotonically until reaching its minimum at  $1/Ro \approx 
 10$, and then it increases monotonically towards the 2D value for a two-roll state. Similar non-monotonic behavior of 
 $Nu$ with the control parameter was also found in sheared RB convection, where $Nu$ also non-monotonically depends on the wall shear Reynolds number \citep{blass2019flow}.
     For span-wise rotating RB convection, $Re$ monotonically increases from the 3D value towards the 2D value with increasing rotation rate, as shown in figure \ref{3d_nu}($b$). Note that this behavior is very different from RB convection  rotating around the  {\it vertical axis}, where 
   $Re$ decreases monotonically with increasing $1/Ro$ \citep{chong2017confined}.

\begin{figure}
 \centering
 \begin{overpic}[width=1\textwidth]{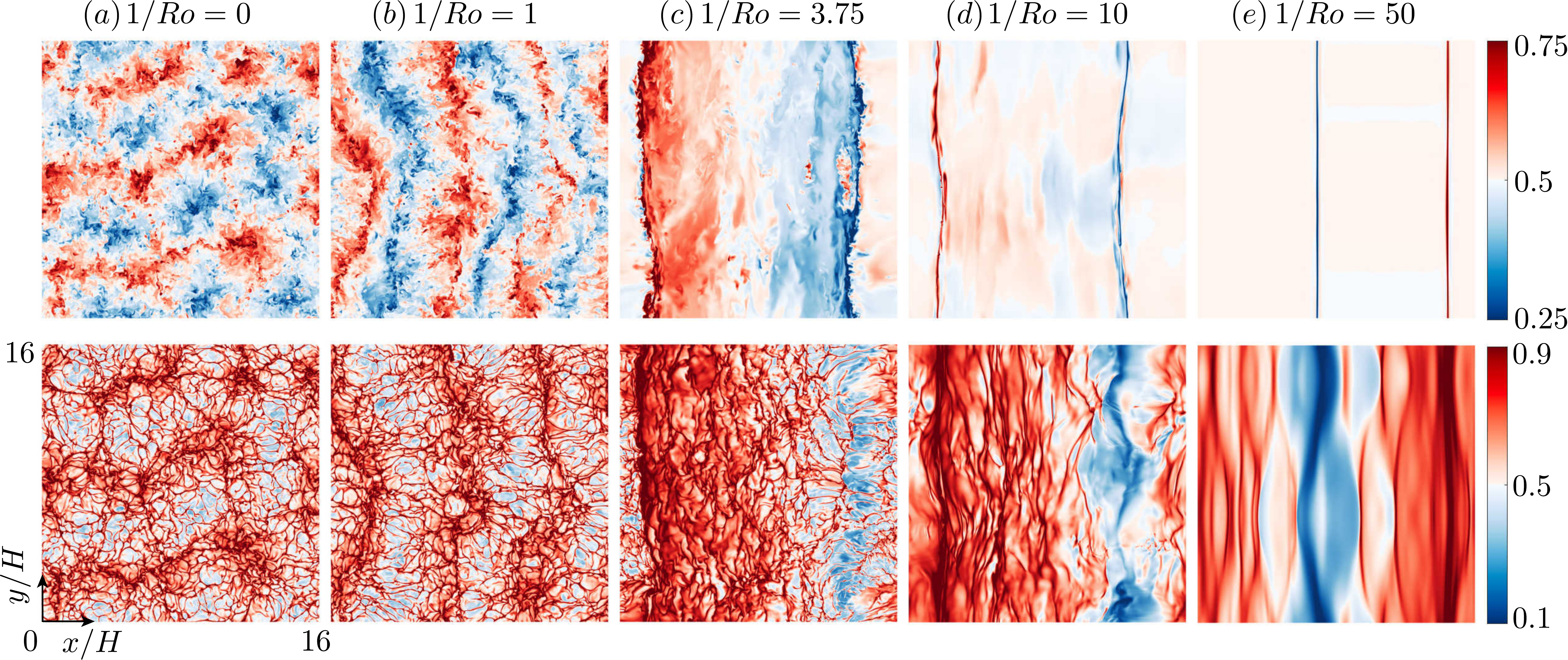}
 \end{overpic} 
 \caption{ 3D RB convection with span-wise rotation (the rotating axis is parallel to $y$ axis): top view 
 snapshots of temperature fields at midheight ($z=H/2$, top row) and thermal boundary layer height ($z= z_{bl} =
 H/ (2 Nu)$, bottom row) for $Ra=10^7,Pr=0.71,\Gamma=16$ with different rotation rates. (\textit{a}) $1/Ro=0$.  (\textit{b}) $1/Ro=1$. (\textit{c}) $1/Ro=3.75$. (\textit{d}) $1/Ro=10$. (\textit{e}) $1/Ro=50$.  }\label{3d_flow}
\end{figure}

We now connect the global transport properties with the flow organization. Figure \ref{3d_flow} shows instantaneous temperature fields at the midheight (top row) and boundary layer height close to the bottom plate (bottom row) for different $1/Ro$ with $Ra=10^7,Pr=0.71,\Gamma=16$. We can clearly see the connection between large-scale thermal structure at midheight and boundary layer height for different $1/Ro$, which has also been shown in 3D RB convection with no-slip plates \citep{stevens2018turbulent,green2020resolved}. With increasing rotation rate, one sees the increasing two-dimensionlization of the flow. For the non-rotation case ($1/Ro=0$), figure \ref{3d_flow}($a$) shows qualitatively similar superstructures as the no-slip case \citep{pandey2018turbulent,stevens2018turbulent}. When $1/Ro$ increases to 1, 
a meandering large-scale convection roll state develops, as can be seen from
 figure \ref{3d_flow}($b$). Interestingly, similar meandering structures have also been observed in many shear-driven flows when the horizontal isotropy is broken, such as  plane Couette flow \citep{lee2018extreme}, wavy Taylor rolls in Taylor-Couette flow \citep{andereck1986flow}, and also in sheared RB convection \citep{blass2019flow}. Figure \ref{3d_nu}($a$) shows that this meandering structure at $1/Ro=1$ has still similar $Nu$ as in the non-rotation case. 
  For medium rotation rates  $1/Ro=3.75$ and 10,  a two-roll state evolves;
   interestingly,  the cyclonic circulation has a larger size than the anticyclonic one. The smaller $Nu$ for these two-roll states is related to the decreased plume emission area. With further increasing rotation rate, the flow increasingly behaves like 2D cases. For the largest $1/Ro=50$ as shown in figure \ref{3d_flow}($e$),  a two-roll state with equal size of each roll has developed,
    with small span-wise variation in temperature.

After we have shown the increasing two-dimensionlization of the flow with increasing rotation rate for span-wise rotating 
RB convection,  next we will study the transition from zonal flow to the convection roll states with increasing aspect ratio $\Gamma$, similarly as we have already done for the 2D case. 
\cite{von2015generation} studied span-wise rotating RB convection  for $Ra=10^7,Pr=0.71$ with fixed $\Gamma=2\pi$. They observe strong zonal flow that is perpendicular to both rotation vector and gravity vector. Both the cyclonic zonal flow and the anticyclonic one have been obtained using different initial conditions. These two kinds
 of zonal flow are symmetric
  for 2D, while they are not in 3D with span-wise rotation, as the 
  Coriolis force depends on the direction of velocity and thus it breaks the symmetry between the two kinds of zonal flow, which have an opposite flow direction. The main difference between the two kinds of zonal flow is that intermittent bursts exist for anticyclonic zonal flow, similar to what is observed in 2D cases with small $Pr\le2$, while these bursts are absent for the cyclonic zonal flow \citep{von2015generation}. We note that, 
 as in \cite{von2015generation},
  the dimensionless angular velocity  $\Omega^\prime=\Omega\tau_{th}$ 
  is used to quantify the rotation velocity, where $\tau_{th}=H^2/\kappa$ is thermal diffusive time. The dimensionless angular 
  velocity is related to the Rossby number  by $Ro=\sqrt{RaPr}/(2\Omega^\prime)$ \citep{novi2019rapidly}.

\begin{figure}
 \centering
 \begin{overpic}[width=0.93\textwidth]{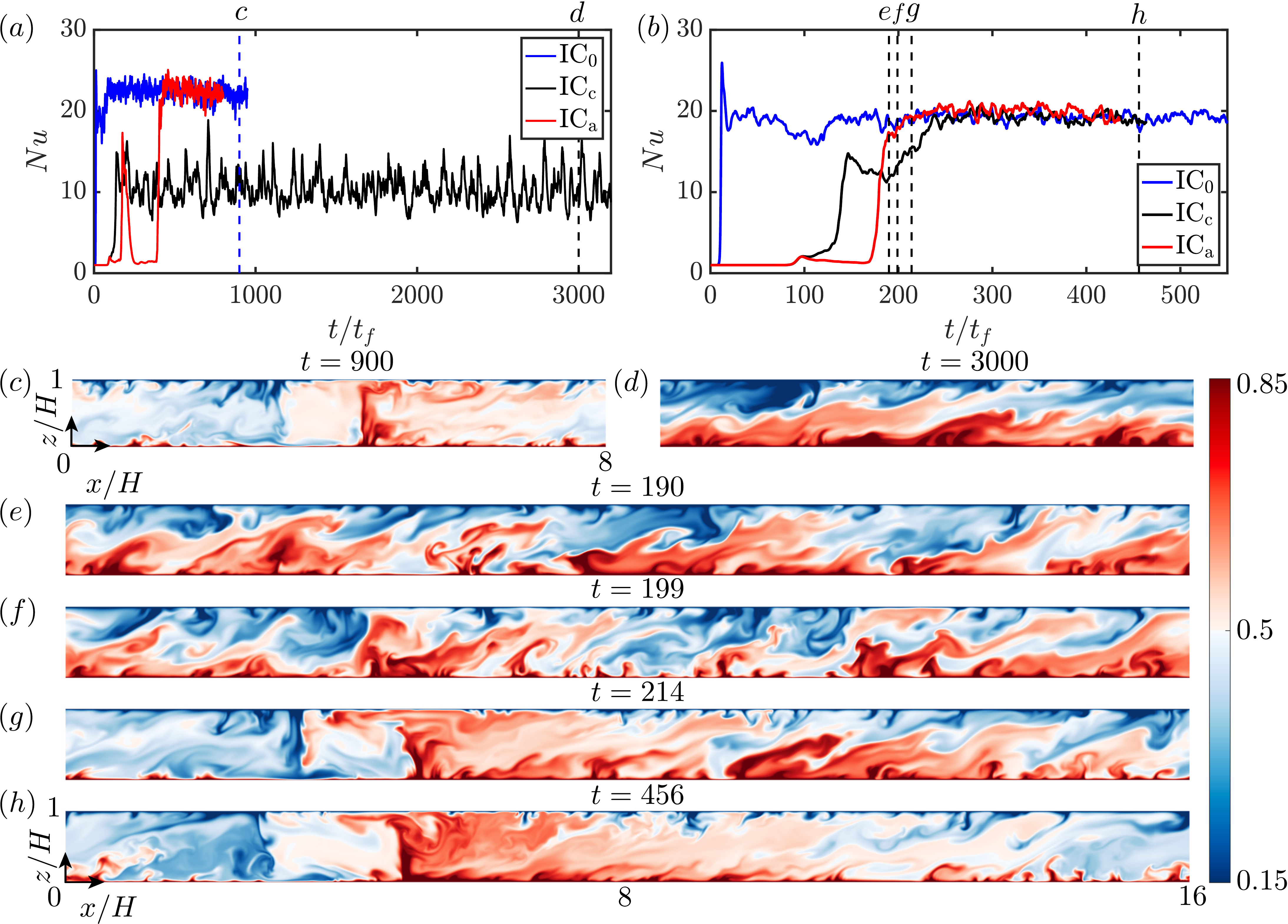}
 \end{overpic} 
 \caption{3D RB convection  with span-wise rotation (the rotating axis is parallel to $y$ axis): time evolution of $Nu$ for $Ra=10^7, Pr=0.71, 1/Ro=3.75$ with three different initial conditions for (\textit{a}) $\Gamma=8$ and (\textit{b}) $\Gamma=16$. $\rm{IC_0}$ means initial condition with zero velocity. $\rm{IC_c}$/ $\rm{IC_a}$  denote cyclonic/anticyclonic shear flow as initial condition. (\textit{c-h}) Side view temperature snapshots at mid span-wise length ($y/H=\Gamma/2$) at different times denoted by the dashed lines in (\textit{a}) and (\textit{b}). }\label{3d_zonal}
\end{figure}

We performed
 simulations for $Ra=10^7$, $Pr=0.71$, $\Gamma=8$ and 16 with $1/Ro=3.75$,  which corresponds to $2\Omega^\prime=10000$ in \cite{von2015generation}. 
 Three different initial conditions were
  used to trigger different possible states, namely:
 \begin{itemize}
 \item 
  $\rm{IC}_0$ with  zero initial velocity, 
  \item  
  $\rm{IC}_c$ with cyclonic shear flow $u(z)=2z-1$, $v=0$, $w=0$, to trigger possible cyclonic  zonal flow, and 
  \item 
  $\rm{IC}_a$ with anticyclonic shear flow $u(z)=1-2z$, $v=0$, $w=0$, to trigger  possible anticyclonic zonal flow. 
  \end{itemize}
 
  We first report the results for $\Gamma = 8$: 
  Figure  \ref{3d_zonal}($a$) illustrates that for  the initial conditions  $\rm{IC}_0$, 
  the flow quickly develops into a two-roll state as indicated in figure \ref{3d_zonal}($c$). 
  We note that the cyclonic circulation is again larger than the anticyclonic one. 
  For $\rm{IC}_c$ as initial conditions, the cyclonic zonal flow shown in  figure  \ref{3d_zonal}($d$) can statistically 
  stably exist for more than 3000 free-fall time units. Finally, the initial conditions 
  $\rm{IC}_a$ can trigger an anticyclonic zonal flow with burst phenomenon, which is consistent with
  the findings of 
   \cite{von2015generation}. However, here this feature 
    only lasts for about 380 free-fall time units and then the system 
    undergoes a transition to a two-roll state. We thus conclude that for $\Gamma=8$ 
    the system can again display a  bi-stability behavior, in which
     both zonal flow and convection roll states are statistically stable.

 We now come to the case of  $\Gamma=16$.
 From figure \ref{3d_zonal}($b$) we conclude that for three different initial conditions, 
  the flow eventually evolves to the very same final state, namely the convection roll state. 
  The cyclonic zonal flow initially
  seen for short time for the initial conditions $\rm{IC}_c$
   quickly undergoes a transition to the convection roll state. 
   The snapshots in figure \ref{3d_zonal}($b-d$) reveal  similar transition processes 
    as we had already observed in 2D. 
   Again, the final two-roll state has a larger cyclonic circulation. For the final convection state the horizontal scale of the flow reaches the domain size of 16, which is much larger than the typical horizontal scale of superstructures observed in 3D RB convection
  with no-slip plates \citep{pandey2018turbulent,stevens2018turbulent}. Such large-scale structures can not be captured in small domains, which is the reason why for small domains
  only zonal flow states can be realized
   \citep{von2015generation}. 
 
 To summarize our results from our 3D simulations with free-slip plates and span-wise rotation 
 (with fixed $Ra=10^7$, $Pr=0.71$, and relative strong rotation $1/Ro=3.75$): We have revealed a
  similar physical picture  as we had done before 
  for 2D RB convection with free-slip plates: 
  \begin{itemize}
  \item[(i)]
  For small aspect ratio $\Gamma=2\pi$, the flow is zonal \citep{von2015generation}.
  \item[(ii)]
    With increasing $\Gamma$ up to at least $\Gamma = 8$, the convection roll state and the zonal flow state
    coexist in phase space
    and which one is taken depends on the initial conditions.
   \item[(iii)]  
    For larger $\Gamma=16$, we always obtain the convection roll state, independently on 
    of what kind of initial conditions were employed. 
\end{itemize}

\section{Concluding remarks} \label{sec5}
In summary, we have studied 2D RB convection  and span-wise rotating RB convection
 with free-slip plates in horizontally periodic domains using direct numerical simulations. We find that the zonal flow, which was previously observed in a $\Gamma=2$ cell \citep{van2014effect,goluskin2014convectively}, cannot be
  sustained and will undergo transitions to convection roll states when the aspect ratio $\Gamma$ is larger than a critical value, which depends on the Rayleigh number $Ra$ and Prandtl number $Pr$. 
  
  We reveal three regimes: 
  \begin{itemize} \item[(i)]
  For small $\Gamma$ (typically $\Gamma\le1-3$, depending on $Ra$ and $Pr$), 
  only zonal flow can be observed. 
  \item[(ii)]
  With increasing $\Gamma$, we first find a bi-stable regime in which, depending on the initial conditions, both zonal flow and convection roll states can be stable .
  \item[(iii)]  For even larger-aspect ratio systems, only convection roll states can be sustained. 
  \end{itemize} 
  How many convection rolls develop in the convection roll states 
  (in other words, what is the mean aspect ratio $\Gamma_r$ of an individual roll) depends on the initial conditions. For instance, 
  for $Ra=10^8$ and $Pr=10$ 
   the horizontal extent of the stable convection rolls varies between $16/11$ and $
   64$ times the height of the convection cell.
  A convection roll with an as large aspect ratio of $\Gamma_r = 64$ or more generally already with $\Gamma_r \gg 10$ can be seen as “localized” zonal flow.

The heat transfer in the system increases significantly when the horizontal extent of the convection roll is reduced. It is 
found that the Prandtl number dependence of the Nusselt number
$Nu(Pr)$ has very different trends for large and small $\Gamma_r$: 
For large $\Gamma_r$ (like $\Gamma_r=16$ for a $\Gamma=32$ cell), the flow behaves  like a  ``localized'' zonal flow state, and $Nu$ increases with increasing $Pr$,  similarly  as we found for the  zonal flow state.
 In contrast,  for small $\Gamma_r$, the heat flux into the system is dominated by the plume-impacting regions on the bottom
  plate,  in which the local heat flux is very high and increases with decreasing $Pr$, implying that 
   the global heat flux $Nu$ also increases with decreasing $Pr$.

For span-wise rotating 3D RB convection, we find that with increasing rotation rate $1/Ro$, 
both the transport properties (like the Nusselt number $Nu$ and the Reynolds number $Re$) and the flow organization
increasingly behave like in the corresponding 2D cases.  
In particular, just as in 2D, the
 zonal flow observed in a small periodic cell  with $\Gamma=2\pi$ \citep{von2015generation},
 disappears in larger  cells with $\Gamma=16$. 
 For intermediate $\Gamma=8$, 
 bistability is observed, again similarly as observed in 2D RB convection. 
 
 Finally, an interesting but still open question is the final fate of the aspect ratio dependence of the zonal flow for higher $Ra$, i.e., is there a finite $Ra$ above which zonal flow exists for all $\Gamma$? Due to the chaotic nature of the flow, mapping out the parameter regime where zonal flow can be found is not easy, especially not for high $Ra$ and large $\Gamma$. 

From a broader perspective, our study underlines the importance of having  
large enough aspect ratios in numerical simulations of wall-bounded turbulent
flows, even when one employs periodic boundary conditions. We had seen this before
 in 3D RB convection with no-slip
velocity boundary conditions at the plates
\citep{pandey2018turbulent,stevens2018turbulent,krug2020coherence,green2020resolved}, but apparently this conclusion is much more general.

\section*{Acknowledgements}
We thank Olga Shishkina and David Goluskin for fruitful discussions. Q.W. acknowledges financial support from China Scholarship Council (CSC) and Natural Science Foundation of China under grant no 11621202.  K.L.C. acknowledges the Croucher Foundation for the Croucher Fellowships for Postdoctoral Research. R.J.A.M.S. acknowledges the financial support from ERC (the European Research Council) Starting Grant No. 804283 UltimateRB. We acknowledge PRACE for awarding us access to MareNostrum 4 based in Spain at the Barcelona Computing Center (BSC) under Prace project 2018194742 and to Marconi based in Italy at CINECA under PRACE project 2019204979. This work was partly carried out on the national e-infrastructure of SURFsara, a subsidiary of SURF cooperation, the collaborative ICT organization for Dutch education and research.

\newpage

\begin{appendix}

\section{Tables with  simulation details }

\tabcolsep 2.pt
\setlength{\LTcapwidth}{1\linewidth}
\renewcommand{\arraystretch}{1}
\begin{longtable}{ccccccccccccc}
\hline
    $Ra$            &    $Pr$          & $\Gamma$     &$N_x\times N_z$  &  $n^{(i)}$          &  $n$  	&$\Gamma_r$         &  $Nu$         & $Re$       & $Re_x$         & $Re_z$              & $t_{tot}$            & $t_{avg}$  \\
\hline
\endfirsthead
\hline
     $Ra$            &    $Pr$          & $\Gamma$     &$N_x\times N_z$	 & $n^{(i)}$        &  $n$ & $\Gamma_r$         &  $Nu$         & $Re$      & $Re_x$         & $Re_z$               & $t_{tot}$            & $t_{avg}$  \\
\hline
\endhead
\hline
\endfoot

$10^7$         &   10              & 16                    & $4096\times256$     & 2       & 2   &8      & 25.67        &   413.19        &  405.27          & 80.52                 & 7000            & 5000 \\
$10^7$         &   10              & 16                    & $4096\times256$     &  4       &4  &4      & 30.95        &448.47           & 427.38           &  135.91                 & 7000            &5000  \\
$10^7$         &   10              & 16                    & $4096\times256$     & 6        & 6  &2.67      & 40.84        &  493.15         &453.37            &194.05                  &  7000           & 5000 \\
$10^7$         &   10              & 16                    & $4096\times256$     & 8        & 8 &2      & 41.94        &  474.28         &417.81            & 224.44                   &   7000          &  4000\\
$10^7$         &   10              & 16                    & $4096\times256$     & 10        & 10  &1.6      & 48.53        &  488.77         &   409.76         & 266.45                 & 7000            & 3000 \\
&                     &                        &                               &                   &                        &                       &                  &                             &               &     &  \\
$3\times10^7$         &   10              & 16      & $6144\times384$  &2        &  2  &8        &36.16         &881.17           &864.21            &172.06                  &  6500           &4500  \\
$3\times10^7$         &   10              & 16       & $6144\times384$    &4         & 4  &4       & 46.15         & 956.10          & 910.58           &291.51                  &  6500           &4500  \\
$3\times10^7$         &   10              & 16       & $6144\times384$  &  6       & 6  &2.67        &  55.61       & 997.30          &   917.21         &  391.60                  & 11500            & 5500 \\
$3\times10^7$         &   10              & 16     & $6144\times384$    & 8        &  8   &2      & 64.98        &1033.58           &910.66            & 488.87                 &  11500           &7500  \\
$3\times10^7$         &   10              & 16     & $6144\times384$   &10         &10    &1.6        & 69.43        &1016.46           & 852.94           & 552.91                 & 6500            & 4500 \\
&                     &                        &                               &                   &                        &                       &                  &                             &               &     &  \\
$10^8$         &   10              & 16                    & $6144\times384$  &  2       & 2   &8  &53.30         & 1995.38          & 1955.07         &399.06                  &6000             &4000  \\
$10^8$         &   10              & 16                    & $6144\times384$   & 4        & 4   &4  &67.66         &2112.39           &  2011.37          &645.40                  & 6000            & 4000 \\
$10^8$         &   10              & 16                    & $6144\times384$   &6         &  6  &2.67 & 86.49        & 2296.92          & 2112.33            & 902.13                 &  8000           & 6000 \\
$10^8$         &   10              & 16                    & $6144\times384$  & 8        &8  &2 & 94.10        & 2281.70          & 2010.51           &  1078.91                   &6000             & 3000 \\
$10^8$         &   10              & 16                    & $6144\times384$    & 10        & 10   &1.6 &96.27         &2200.71           & 1846.23           & 1197.73                 &  6000           & 3000 \\
$10^8$         &   10              & 16                    & $6144\times384$   &12         &  10   &1.6  &96.20         & 2198.65          &1844.19            &1197.09                & 6000            &3000 \\
$10^8$         &   10              & 16                    & $6144\times384$    &  14  & 10   &1.6 & 96.81        &  2202.08         & 1846.89           &1199.23                  & 2456            & 1456 \\
$10^8$         &   10              & 16                    & $6144\times384$    &  16  & 10   &1.6 & 96.20        & 2200.23        & 1845.68           &1197.70                & 12000            & 10000 \\
&                     &                        &                               &                   &                        &                       &                  &                             &               &     &  \\
$3\times10^8$         &   10              & 16       & $8192\times512$    &2         & 2   &8      & 75.85        &  4154.56         & 4068.12           &  843.14                & 5500            & 3500 \\
$3\times10^8$         &   10              & 16       & $8192\times512$   & 4        &   4  &4       &  92.95       & 4298.72          &  4092.94          &  1314.09               & 5500            & 3500 \\
$3\times10^8$         &   10              & 16        & $8192\times512$  &6         & 6   &2.67        &  121.79       &  4738.24         & 4357.01           &1862.08                   &5500             &3500  \\
$3\times10^8$         &   10              & 16       & $8192\times512$   &8         &  8  &2       &129.09         & 4675.29          &  4121.31          & 2207.51                  &5500             & 3500 \\
$3\times10^8$         &   10              & 16       & $8192\times512$    & 10        & 10  &1.6       & 134.00        & 4552.41          &  3821.76          &2473.57                  & 5500            &3500  \\
&                     &                        &                               &                   &                        &                       &                  &                             &               &     &  \\
$10^9$         &   10              & 16                    & $12288\times768$   &2         & 2  &8         & 112.36        & 9258.76          & 9061.98           & 1898.78                  &   1600          & 1100 \\
$10^9$         &   10              & 16                    & $12288\times768$  & 4        & 4   &4        & 137.80        & 9622.01          &9160.21            & 2945.08                   & 5300            &  1300\\
$10^9$         &   10              & 16                    & $12288\times768$   &  6       &  6  & 2.67       & 171.87        & 10380.02          & 9548.89           & 4069.56                 &   5046          &2046  \\
$10^9$         &   10              & 16                    & $12288\times768$   &  8       &   8  &2       &  186.55       & 10373.03          &9148.56            & 4889.31                & 4742            &  2742\\
$10^9$         &   10              & 16                    & $12288\times768$   & 10      & 10 &1.6        & 198.01        &10135.10           &8504.77            & 5512.29                  &3700             & 1700 \\
&                     &                        &                               &                   &                        &                       &                  &                             &               &     &  \\
  \caption{Simulation details for all cases shown 
  in figure \ref{pr10_roll}($a$).  The columns from left to right indicate $Ra$, $Pr$,  $\Gamma$, grid resolutions $N_x\times N_z$, the number of initial rolls $n^{(i)}$, the number of final convection rolls $n$, the mean aspect ratio of the convection rolls $\Gamma_r=\Gamma/n$, the Nusselt number $Nu$,  the Reynolds number $Re$ based on root-mean-square of the global velocity, the horizontal Reynolds number $Re_x$ based on root-mean-square of the horizontal velocity, the
  vertical Reynolds number $Re_z$, the total simulation time $t_{tot}$,
   and the time $t_{avg}$ used to average $Nu$ and $Re$. }
    \label{tabra}  \\ 

 \end{longtable}
\clearpage

\tabcolsep 3pt
\setlength{\LTcapwidth}{1.0\linewidth}
\renewcommand{\arraystretch}{1.0}
\begin{longtable}{ccccccccccccc}
 
\hline
    $Ra$            &    $Pr$          & $\Gamma$     &$N_x\times N_z$  &  $n^{(i)}$          &  $n$  	&$\Gamma_r$         &  $Nu$         & $Re$       & $Re_x$         & $Re_z$              & $t_{tot}$            & $t_{avg}$  \\
\hline
\endfirsthead
\hline
     $Ra$            &    $Pr$          & $\Gamma$     &$N_x\times N_z$	 & $n^{(i)}$          &  $n$ & $\Gamma_r$         &  $Nu$         & $Re$      & $Re_x$         & $Re_z$               & $t_{tot}$            & $t_{avg}$  \\
\hline
\endhead
\hline
\endfoot

$10^8$         &   1              & 16                    & $6144\times384$   & 2           &   2  &8       & 55.04        &   20633.77        &  20190.00          & 4261.16                        &   4000               &   s   \\
$10^8$         &   1              & 16                    &   $6144\times384$   &   4     & 4  &4         &  81.63        &24070.00           & 22920.00        & 7357.00                   &6000   &s   \\
 &                     &                        &                               &                   &                        &                       &                  &                             &               &     &  \\
$10^8$         &   3              & 16                    & $6144\times384$  &2         & 2  &8  & 52.18        &  6679.56         &  6536.21          & 1376.25                    &10000             & 2000 \\
$10^8$         &   3              & 16                    & $6144\times384$ &4         & 4   &4  & 78.27        & 7822.00          & 7448.00           &  2391.00                  & 5434            & 1434 \\
$10^8$         &   3              & 16                    & $6144\times384$    &  6       & 6   &2.67 & 93.06        &8156.64           & 7502.60           &3200.30                  & 5345            &2345  \\
$10^8$         &   3              & 16                    & $6144\times384$  &  8       &  8  &2  &102.51       & 8188.54          &  7220.01          &  3863.14                &   4600          &1600  \\
&                     &                        &                               &                   &                        &                       &                  &                             &               &     &  \\
%
$10^8$         &   30              & 16                    & $6144\times384$   &   2      & 2  &8 & 56.69        &646.20           &634.14            &124.28                    &6000             & 4000 \\
$10^8$         &   30              & 16                    & $6144\times384$   & 4        &4   &4 & 64.17        &  670.83         & 639.73           &  201.89                 &  6000           & 4000 \\
$10^8$         &   30              & 16                    & $6144\times384$   & 6        &6   &2.67  & 77.54        &697.39           &641.45            & 273.67                  &4311           &3311  \\
$10^8$         &   30              & 16                    & $6144\times384$   &  8       & 8  &2 & 84.60        & 688.69          & 606.92           & 325.50                 & 6000            &  4000\\
$10^8$         &   30              & 16                    & $6144\times384$  &  10       &  10  &1.6 & 89.96        &673.49           & 565.05           &   366.48               &  6000           &4000  \\
$10^8$         &   30              & 16                    & $6144\times384$   & 12        &   10  &1.6       &90.13         &  674.42         & 565.86           & 366.94                 &  3000     & 2000 \\
$10^8$         &   30              & 16                    & $6144\times384$  &  14       &   10 &1.6        & 89.99        & 673.58          & 565.10           &   366.58               &4000             &3000  \\
$10^8$         &   30              & 16                    & $6144\times384$   &  16       & 10   &1.6       & 89.94        &673.42           &565.20            & 366.13                 &  4000           & 3000 \\
&                     &                        &                               &                   &                        &                       &                  &                             &               &     &  \\
$10^8$         &   100              & 16                    & $6144\times384$  &    2     &   2  &8   & 57.98        &163.70           & 160.24           &33.45                & 5000            & 2000 \\
$10^8$         &   100              & 16                    & $6144\times384$   &   4      &4   &4 & 62.53        & 179.11          & 171.18           &   52.71                &  6000           &5000  \\
$10^8$         &   100              & 16                    & $6144\times384$   &  6       &  6 &2.67 &68.65         & 187.57          & 172.86           &  72.82                 &    6000         & 5000 \\
$10^8$         &   100              & 16                    & $6144\times384$  &   8      & 8   &2 &74.72         &184.90           & 163.07           &  87.16                & 6000            &5000  \\
&                     &                        &                               &                   &                        &                       &                  &                             &               &     &  \\

$10^8$         &   1              & 32                    & $12288\times384$   &  2       & 2 &16        &35.69         &  16990.00         & 16820.00           & 2421.00                   & 6000            &s  \\
$10^8$         &   1              & 32                    & $12288\times384$   &4         & 4   & 8       & 55.06        &20656.18           & 20210.0           & 4264.00                  &  5000           & s \\
$10^8$         &   1              & 32                    & $12288\times384$     & 6        &  6 &5.33       & 69.79        & 22760.00          &21989.82            & 5886.63                 & 5278            &s  \\
$10^8$         &   1              & 32                    & $12288\times384$     &  8       & 8 &4     & 81.62        & 24070.44          &22918.82            & 7356.22                   &   4177          &s  \\

&                     &                        &                               &                   &                        &                       &                  &                             &               &     &  \\
$10^8$         &   3              & 32                    & $12288\times384$     &   2      &  2  &16     &44.84         &  6337.77         &6274.55            &  892.89                &6500             &4500  \\
$10^8$         &   3              & 32                    & $12288\times384$    & 4       & 4    &8      &51.70         & 6647.40          &  6504.83          &  1369.35                & 6500            &  4500\\
$10^8$         &   3              & 32                    & $12288\times384$    & 6        &  6   &5.33      & 69.52        & 7526.00          &7270.00            & 1947.00                & 3678            &s  \\
$10^8$         &   3              & 32                    & $12288\times384$    &  8       & 8 &4      &  78.26       & 7820.16          &  7446.00          & 2390.36                   & 4032            &2032  \\
$10^8$         &   3              & 32                    & $12288\times384$    & 10        & 10  &3.2      &84.72         &7960.42           &  7456.85          & 2786.39                   & 3440            &  1440\\
$10^8$         &   3              & 32                    & $12288\times384$     & 12        & 12  &2.67     & 93.09        &8154.74           & 7500.63           & 3200.02                   & 5205            & 3205 \\
$10^8$         &   3              & 32                    & $12288\times384$      & 14        & 14  &2     &99.91        &  8236.24         & 7422.50           &3569.63                  &3738             & 1738 \\
&                     &                        &                               &                   &                        &                       &                  &                             &               &     &  \\
$10^8$         &   10              & 32                    & $12288\times384$    & 2        &2  &16       &52.96         & 2037.02          & 2017.49           &  281.43                  &   6000          &4000  \\
$10^8$         &   10              & 32                    & $12288\times384$       &  4       &4  &16     &53.25         & 1994.84          & 1954.55           &  398.86                 & 6000            &4000  \\
$10^8$         &   10              & 32                    & $12288\times384$    & 6       &6 & 5.33       &62.34         &  2090.82         &2020.23            &  538.71                  &  6000           & 4000 \\
$10^8$         &   10              & 32                    & $12288\times384$     &8         & 8 &4     & 67.62        &2111.59           &2010.61            &645.18                    & 6000            & 4000 \\
$10^8$         &   10              & 32                    & $12288\times384$     &  10       & 10  &3.2      & 71.35        &  2121.42         &1987.41           & 742.14                 & 6000            &4000  \\
$10^8$         &   10              & 32                    & $12288\times384$    &  12       &  12  &2.67      & 87.73        & 2313.32          &2127.41            & 908.67                 &  6000           &4000  \\
$10^8$         &   10              & 32                    & $12288\times384$      &14         & 14  &2.29     & 91.47        & 2307.20          &2078.88            &1000.70                  & 6000            & 3000 \\
$10^8$         &   10              & 32                    & $12288\times384$   & 16        &   16 &2       & 95.01        &2291.69           & 2019.13           & 1083.97                  &6000             & 3000 \\
$10^8$         &   10              & 32                    & $12288\times384$     &18        &    18  &1.78       & 97.25        &  2261.39         &  1945.47          & 1152.82              & 6000           &3000  \\
$10^8$         &   10              & 32                    & $12288\times384$    &  20      & 20  &1.6       & 96.13        & 2199.85          &  1845.37          &1197.48                   &  6000           &3000 \\
$10^8$         &   10              & 32                    & $12288\times384$     &22         &22 &1.45      &  97.22       &  2155.22         & 1761.17           &  1242.27                   &  6000           & 3000 \\
$10^8$         &   10              & 32                    & $12288\times384$    &  24       &   22  &1.45      & 97.53        &  2158.19         &  1763.30          &  1244.42                &  2587           &  1087\\
&                     &                        &                               &                   &                        &                       &                  &                             &               &     &  \\
$10^8$         &   30              & 32                    & $12288\times384$  &  2          & 2  &16        & 56.36        &  670.65         &  664.57          & 90.07          &5800         & 4000          \\
$10^8$         &   30              & 32                    & $12288\times384$   &  4           &4   &8       & 56.76        & 646.62          &634.56            & 124.33         &  6000       & 4000            \\
$10^8$         &   30              & 32                    & $12288\times384$   &  6           &  6 &5.33       & 59.03        &  651.68         & 631.49           & 160.97          &  5800       & 4000           \\
$10^8$         &   30              & 32                    & $12288\times384$   & 8            & 8 &4         &  64.22       &671.01           &639.90            &201.94           & 6000        & 4000           \\
$10^8$         &   30              & 32                    & $12288\times384$    &  10           & 10 &3.2       & 70.29        &  683.07         & 640.19           &238.19           & 6000        & 4000          \\
$10^8$         &   30              & 32                    & $12288\times384$  & 12           &12 &2.67        & 77.55        & 697.21          &641.28            & 273.60          &   6000      &  4000           \\
$10^8$         &   30              & 32                    & $12288\times384$   & 14            & 14 &2.29        & 81.51        &695.09           & 626.40           & 301.29          & 6000        & 4000         \\
$10^8$         &   30              & 32                    & $12288\times384$   &  16           &  16 &2        & 84.53        &688.59           & 606.83           &  325.45         &  6000       & 4000         \\
$10^8$         &   30              & 32                    & $12288\times384$  &  18           &18  &1.78        & 88.06        & 684.88          &589.31            &348.96           &  6000       &  4000           \\
$10^8$         &   30              & 32                    & $12288\times384$    &   20         & 20  &1.6       &89.93         &673.55           & 565.16           & 366.41          &  6000       &  4000        \\
&                     &                        &                               &                   &                        &                       &                  &                             &               &     &  \\

$10^8$         &   100              & 32                    & $12288\times384$ & 2            &2  &16         &56.69         &161.43           &159.12            &27.17           & 5500        &   3500         \\
$10^8$         &   100              & 32                    & $12288\times384$ &  4           & 4  &8         & 58.09        &  163.88         & 160.43           &  33.47         & 6000        & 4000           \\
$10^8$         &   100              & 32                    & $12288\times384$  &  6           &6 &5.33        &  60.23       & 173.48          & 167.99           &43.28           &  5800       &  4000          \\
$10^8$         &   100              & 32                    & $12288\times384$  &  8           & 8 &4        & 62.53        &178.89           & 170.97           & 52.62          & 6000        &   4000         \\
$10^8$         &   100              & 32                    & $12288\times384$ & 10            & 10 &3.2         &65.79         &186.90           &175.65            & 63.88          & 6000        & 4000           \\
$10^8$         &   100              & 32                    & $12288\times384$  & 12            & 12 &2.67        &  68.81       &187.90           &173.15            &72.98           &6000         &   4000         \\
$10^8$         &   100              & 32                    & $12288\times384$   &  14           & 14 &2.29       & 71.83        &186.70           &  168.44          & 80.53          & 6000        &   4000        \\
&                     &                        &                               &                   &                        &                       &                  &                             &               &     &  \\
   \caption{Simulation details for the main cases in figure \ref{pr10_roll}($b$) for $\Gamma=16$ and 32.  The table head is the same as table \ref{tabra}. ``s'' denotes that the final flow state is steady, which means that both $Nu$ and $Re$ are independent on time once the final state is achieved.}
    \label{tabpr}  \\ 

  \end{longtable}

\clearpage

\tabcolsep 1.8pt
\setlength{\LTcapwidth}{1.0\linewidth}
\renewcommand{\arraystretch}{1.0}
\begin{longtable}{ccccccccccccc}
\hline
$Ra$      & $Pr$    & $1/Ro$  &  $\Gamma$     & $N_x\times N_y\times N_z$    &    IC              &   $Nu$    & $Re$    & $Re_x$     & $Re_y$    &$Re_z$ &  $t_{tot}$    & $t_{avg}$ \\
\hline
\endfirsthead
\hline
    $Ra$      & $Pr$    & $1/Ro$  &  $\Gamma$     & $N_x\times N_y\times N_z$    &    IC              &   $Nu$   & $Re$     & $Re_x$     & $Re_y$    &$Re_z$&  $t_{tot}$       & $t_{avg}$ \\
\hline
\endhead
\hline
\endfoot
$10^7$   & 0.71    &  3.75         &    8                                & $1024\times1024\times128 $   & $\rm{IC}_0$  & 22.32    & 2282.55   & 2052.21     & 641.90 &  765.78 & 950    &  550      \\
$10^7$   & 0.71    &  3.75         &    8                                & $1024\times1024\times128 $   & $\rm{IC}_c$  & 10.32    & 1994.96   & 1929.91     & 424.65 & 273.82   &3200  &  2700      \\
$$   &    &          &                                 &    &  &       &    &      & &       &        \\

$10^7$   & 0.71    &  0         &    16                                & $2048\times2048\times128 $   & $\rm{IC}_0$  & 25.39      & 1184.12   & 707.09     & 733.01 &  604.04  &700   &  400      \\
$10^7$   & 0.71    &  0.1        &    16                                & $2048\times2048\times128 $   & $\rm{IC}_0$  & 25.42      & 1180.84   &702.13      &728.91 & 608.35   &500   & 300       \\
$10^7$   & 0.71    &  0.3         &    16                                & $2048\times2048\times128 $   & $\rm{IC}_0$  &25.66       & 1195.08   &   756.55   & 677.92& 629.49   &500   & 300       \\
$10^7$   & 0.71    &  1         &    16                                & $2048\times2048\times128 $   & $\rm{IC}_0$  & 25.20       & 1409.11   &  1030.76    &712.38  &644.72    &700   & 300       \\
$10^7$   & 0.71    &  3.75    &    16                              & $2048\times2048\times128 $   & $\rm{IC}_0$  &  19.19    &2395.52   &  2217.79   & 680.81      & 596.99    &550   &250        \\
$10^7$   & 0.71    &  10         &    16                                & $2048\times2048\times128 $   & $\rm{IC}_0$  & 18.98      & 3357.81   &3184.53      &741.39 &764.23    &600   & 300       \\
$10^7$   & 0.71    &  20         &    16                             & $2048\times2048\times128 $   & $\rm{IC}_0$  & 19.97      & 3713.41   &3527.17      &784.23  &  856.41   &600  &   200     \\
$10^7$   & 0.71    &  50         &    16                                & $2048\times2048\times128 $   & $\rm{IC}_0$  &23.44       &5436.73    & 5283.16     & 532.86 & 1167.17   &1200   &600        \\
$$   &    &          &                                 &    &  &       &    &      & &       &        \\
$10^7$   & 0.71    &  -(2D)         &    16                             & $4096\times256 $   & $n^{(i)}=2$    &  25.26    &  6270.74  &  6137.35  & -     & 1286.55   &3700  & s       \\
$10^7$   & 0.71    &  -(2D)         &    16                             & $4096\times256 $   & $n^{(i)}=4$   & 37.57    &  7326.16  & 6977.15  & -     & 2234.26    & 3700 & s       \\
$10^7$   & 0.71    &  -(2D)         &    16                             & $4096\times256 $   & $n^{(i)}=6$    &  45.67   &  7702.85  &  7086.30  & -     & 3019.64  &3700   & s       \\
$10^7$   & 0.71    &  -(2D)         &    16                             & $4096\times256 $   & $n^{(i)}=8$    & 51.02    &  7726.20  & 6809.66  & -     & 3650.03  &3700  & s       \\

\caption{Simulations details for 3D RB convection with span-wise rotation. The corresponding 2D simulations are also included for comparison. The table head is similar to that of table \ref{tabra} and \ref{tabpr}, apart from the inverse Rossby number $1/Ro$ for the 3D cases. $\rm{IC_0}$ means initial conditions with zero velocity and conductive temperature profile with random perturbations. $\rm{IC_c}$ means initial conditions with cyclonic shear flow $u(z)=2z-1$, $v=0$, $w=0$ and conductive temperature profile with random perturbations.}
 \label{tab3d}  \\ 

  \end{longtable}

\clearpage

\section{Reynolds number ratio }

\begin{figure}
 \centering
 \begin{overpic}[width=0.9\textwidth]{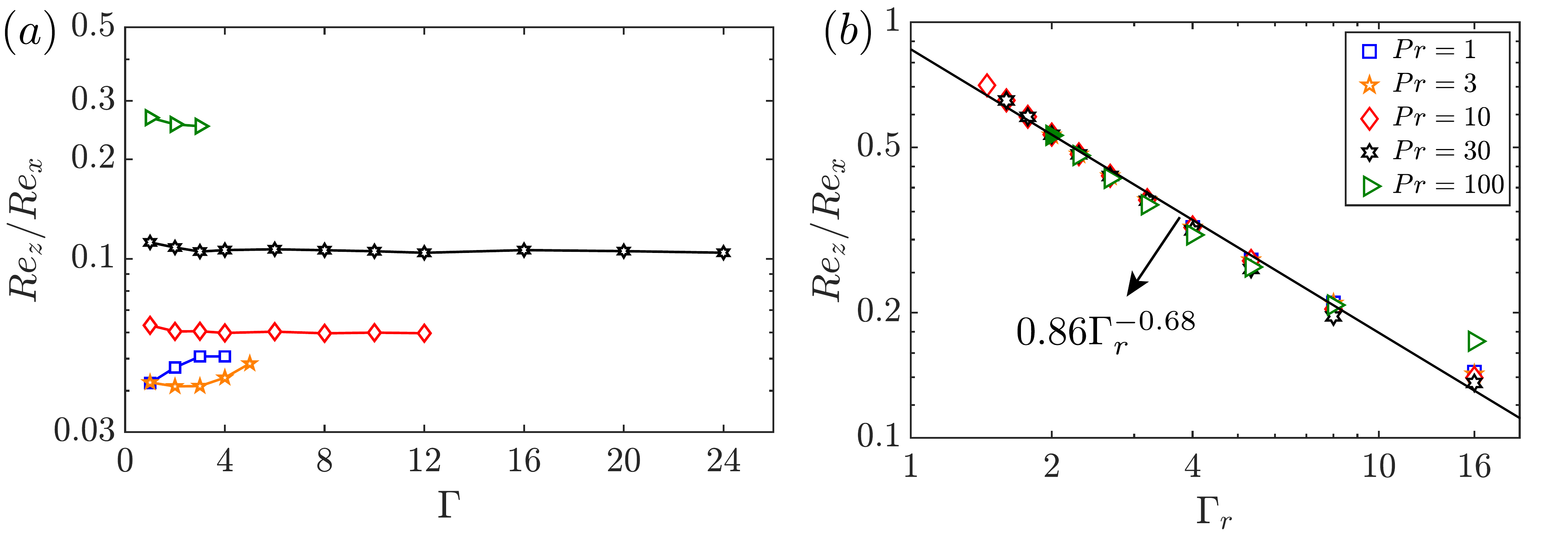}
 \end{overpic}
 \caption{(\textit{a})  Reynolds number ratio $Re_z/Re_x$ as function of  $\Gamma$ for the
 zonal flow state and (\textit{b})  Reynolds number ratio $Re_z/Re_x$  as function of $\Gamma_r$ for the 
 convection roll state for $Ra=10^8$ with different $Pr$. Solid symbols in (\textit{b}) for $\Gamma=16$ and hollow symbols for $\Gamma=32$. The solid symbols can hardly be seen as they are mostly hidden by the hollow symbols. The data in (\textit{b}) can be well described by the  effective scaling relation
  $Re_z/Re_x=0.86\Gamma_r^{-0.68}$. }\label{re_pr}
\end{figure}

Figure \ref{re_pr}($a$) shows the 
Reynolds number ratio $Re_z/Re_x$  as function of  $\Gamma$ for the zonal flow state. 
The ratio $Re_z/Re_x$ increases with increasing $Pr$ for $Pr\ge10$, which has a similar trend as $Nu$ discussed before. Figure \ref{re_pr}($b$) shows Reynolds number ratio $Re_z/Re_x$  as function of $\Gamma_r$ for convection roll states for $Ra=10^8$ with different $Pr$.  It is remarkable 
that $Re_z/Re_x$ seems to have a universal dependence on $\Gamma_r$ for different $Pr$, despite that $Nu$ has a
different trend on $Pr$ for large and small $\Gamma_r$. The data can be well represented by 
the  effective scaling relation  $ Re_z/Re_x =  0.86\Gamma_r^{-0.68}$.

\clearpage

\end{appendix}

\bibliographystyle{jfm}
\bibliography{zonal}

\end{document}